\documentclass[useAMS,usenatbib]{mn2e}

\usepackage{tabularx}
\usepackage{graphicx}
\usepackage{amsmath}
\usepackage{natbib}
\usepackage{lscape}

\title[Overdense environments around \textit{WISE}-selected AGNs]{Overdensities of SMGs around \textit{WISE}-selected, ultra-luminous, high-redshift AGN}

\author[Suzy F. Jones et al.]
{\parbox{\textwidth}{Suzy F. Jones,$^{1}$\thanks{E-mail: suzy.jones@chalmers.se}
Andrew W. Blain,$^{2}$
Roberto J. Assef,$^{3}$
Peter Eisenhardt,$^{4}$
Carol Lonsdale,$^{5}$
James Condon,$^{5}$
Duncan Farrah,$^{6}$
Chao-Wei Tsai,$^{4,7}$
Carrie Bridge,$^{4}$
Jingwen Wu,$^{8}$
Edward L. Wright$^{7}$ and
Tom Jarrett$^{9}$
}\vspace{0.4cm}\\
\parbox{\textwidth}{$^{1}$Department of Space, Earth, and Environment, Chalmers University of Technology, Onsala Space Observatory, SE-43992, Onsala, Sweden\\
$^{2}$University of Leicester, XROA, Department of Physics \& Astronomy, University Road, Leicester LE1 7RH, UK\\
$^3$N\'ucleo de Astronom\'ia de la Facultad de Ingenier\'ia, Universidad Diego Portales, Av. Ej\'ercito Libertador 441, Santiago, Chile\\
$^4$Jet Propulsion Laboratory, California Institute of Technology, 4800 Oak Grove Dr., Pasadena, CA 91109, USA\\
$^5$National Radio Astronomy Observatory, 520 Edgemont Road, Charlottesville, VA 22903-2475 USA\\
$^6$Virginia Polytechnic Institute \& State University, Department of Physics MC 0435, 850 West Campus Drive, Blacksburg, VA 24061, USA\\
$^7$Department of Physics and Astronomy, University of California, Los Angeles, 430 Portola Plaza, Los Angeles, CA, 90095-1547, USA\\
$^{8}$National Astronomical Observatories, Chinese Academy of Sciences, 20A Datun Road, Chaoyang District, Beijing, 100012, China\\
$^{9}$Astronomy Department, University of Cape Town, Rondebosch 7701, Republic of South Africa\\
}}

\begin{document}

\date{Submitted xx/xx/2017}

\pagerange{\pageref{firstpage}--\pageref{lastpage}} \pubyear{2017}

\maketitle

\label{firstpage}

\begin{abstract}
We investigate extremely luminous dusty galaxies in the environments around \textit{WISE}-selected hot dust obscured galaxies (Hot DOGs) and \textit{WISE}/radio-selected active galactic nuclei (AGNs) at average redshifts of $z = 2.7$ and $z = 1.7$, respectively. Previous observations have detected overdensities of companion submillimetre-selected sources around 10 Hot DOGs and 30 \textit{WISE}/radio AGNs, with overdensities of $\sim 2 - 3$ and $\sim 5 - 6 $, respectively. We find that the space densities in both samples to be overdense compared to normal star-forming galaxies and submillimetre galaxies (SMGs) in the SCUBA-2 Cosmology Legacy Survey (S2CLS). Both samples of companion sources have consistent mid-IR colours and mid-IR to submm ratios as SMGs. The brighter population around \textit{WISE}/radio AGNs could be responsible for the higher overdensity reported. We also find the star formation rate density (SFRDs) are higher than the field, but consistent with clusters of dusty galaxies. \textit{WISE}-selected AGNs appear to be good signposts for protoclusters at high redshift on arcmin scales. The results reported here provide an upper limit to the strength of angular clustering using the two-point correlation function. Monte Carlo simulations show no angular correlation, which could indicate protoclusters on scales larger than the SCUBA-2 1.5\,arcmin scale maps.
\vspace{0.6cm}
%Even though there is an overdensity of SMGs, there appears to be no angular clustering of the companion SMGs on the JCMT SCUBA-2 field size ($\sim$ 1.5 arcmin radius). However, the results are not sensitive enough for angular clustering by means of angular two-point correlation function $\omega(\theta)$. The Hot DOGs and \textit{WISE}/radio AGNs are very luminous, heavily obscured, AGN-dominated galaxies. Due to the K-correction effect the SCUBA-2 detection probability should be independent of redshift. We conclude that WISE-selected radio-detected AGN reside in higher overdensities than WISE-selected radio-quiet AGNs. Indeed, this difference is statistically significant even when the different selection effects are taken into account. Therefore, \textit{WISE}/radio AGNs appear to be signposts to overdense pre-virialised environments of galaxies. in 1.5-arcmin radius JCMT SCUBA-2 maps 
%The SMGs in the denser regions around \textit{WISE}/radio AGNs have slightly higher star formation rates by a factor of $\sim$ 18\% compared to SMGs in lower density regions around Hot DOGs.

\end{abstract} 
%by looking at their \textit{WISE} colour-colour plots, mid-IR, submm and radio emission The \textit{WISE} colours and mid-IR to submm ratios of the companion sources are consistent with submillimeter galaxies (SMGs). which could be due to \textit{WISE}/radio AGNs observed with SCUBA-2 having a typical lower redshift, $z = 1.7$, compared to Hot DOGs observed with SCUBA-2 with a typical redshift of $z = 2.7$. Alternatively, \textit{WISE}-selected be due higher radio emission of the \textit{WISE}/radio AGNs compared to Hot DOGs. This could his could be a redshift effect where the Hot DOGs have a lower redshift than the \textit{WISE}/radio AGNs. Alternatively

\section{Introduction}

Advances in infrared (IR) telescope technology like the NASA's \textit{Wide-Field Infrared Survey Explorer} (\textit{WISE}; Wright et al. 2010) have enabled observations of luminous AGN that have been difficult to find with previous IR missions. \textit{WISE} is able to find luminous, dusty, high-redshift, active galaxies because the hot dust heated by AGN and/or starburst activity can be traced using the \textit{WISE} 12\,$\mu$m (W3) and 22\,$\mu$m (W4) bands. \citet{eisenhardt12}, \citet{bridge13} and \citet{lonsdale15} have shown that \textit{WISE} can find different classes of interesting, luminous, high-redshift, dust-obscured AGN. 

%

%\textit{WISE} surveyed the entire sky at wavelengths of 3.4, 4.6, 12 and 22\,$\mu$m (W1-W4) from January 2010 to January 2011 \citep{wright10}. One of the primary science goals was to identify the most luminous galaxies in the observable universe, which can be accomplished due to \textit{WISE} obtaining much greater sensitivity than previous all-sky infrared (IR) survey missions. \textit{WISE} achieved 5-$\sigma$ source sensitivities better than 0.054, 0.071, 0.73 and 5.0\,mJy and angular resolutions of 6.1, 6.4, 6.5 and 12.0\,arcsec in the W1 to W4 bands, respectively \citep{wright10,jarrett11}. The objects observed here are selected from the \textit{WISE} AllWISE Source catalog\footnote{http://wise2.ipac.caltech.edu/docs/release/allwise/}, with IR magnitudes derived using point source profile-fitting \citep{cutri12}. 
% For example, the \textit{Infrared Astronomical Satellite} (\textit{IRAS}) yielded catalogued source sensitivities of 0.5\,Jy at 12, 25 and 60\,$\mu$m and 1\,Jy at 100\,$\mu$m and angular resolutions that varied from 0.5\,arcmin at 12\,$\mu$m to about 2\,arcmin at 100\,$\mu$m \citep{neugebauer84}; compared to 

\citet{eisenhardt12} and \citet{wu12} observed galaxies with faint or undetectable flux densities in the 3.4\,$\mu$m (W1) and 4.6\,$\mu$m (W2) bands, and well detected fluxes in the W3 and/or W4 bands, with a radio blind selection, giving a ``W1W2-dropout'' selection yielding hot, dust obscured galaxies (Hot DOGs). 
%There has been previous work on heavily-obscured, hyper-luminous, \textit{WISE}-selected AGNs from 

Another population of  luminous, dusty, \textit{WISE}-selected AGNs were found by \citet{lonsdale15}, by combining \textit{WISE} and National Radio Astronomy Observatory (NRAO) Very Large Array (VLA) Sky Survey (NVSS) \citep{condon98} and/or Faint Images of the Radio Sky at Twenty-cm (FIRST) \citep{becker95}. They were selected in a similar method in the mid-IR, and are a similarly high luminosity, dust-obscured population and in this paper are known as \textit{WISE}/radio AGNs. The strong compact radio emission could be from AGN jets \citep{lonsdale15}.

%The \textit{WISE}/radio AGNs had more relaxed mid-IR selection cuts and potential bias caused by polycyclic aromatic hydrocarbon (PAH) emission/silicate features that could be redshifted into the mid-IR SED (Lonsdale et al. 2015).

A sample of 10 Hot DOGs and 30 \textit{WISE}/radio AGNs were observed with James Clerk Maxwell Telescope (JCMT) Submillimetre Common-User Bolometer Array 2 (SCUBA-2), and the fields around them were found to have an overdensity of submillimetre galaxies (SMGs)\footnote{Submm galaxies (SMGs) were historically defined by having a submm flux density of $S_{850\mu \rm{m}}$ $> 2$\,mJy. SMGs are massive gas-rich, high-redshift galaxies with high IR luminosities, L$_{\rm{IR}}$ $\ge$ 10$^{12}$ L$_{\odot}$, believed to be from starburst activity, with star formation rates (SFRs) of several 100-1000\,M$_{\odot}$ yr$^{-1}$ \citep{smail97,ivison98,eales99,smail00,blain02,pope06,casey14,swinbank14}.  SMGs are enshrouded by dust and hence are faint in optical and near-IR wavelengths.} by a factor of $\sim$ 2.4 and $\sim$ 5.6, respectively, when compared with blank-field submm surveys in \citet{jones14,jones15}. The Hot DOGs appeared to have redder mid-IR colours and less submm emission than \textit{WISE}/radio AGNs, which could be due to selection effects \citep{jones14,jones15}.  They have very similar SEDs and are both redder than standard AGN templates (see Figure 5 in both Jones et al. 2014 and 2015). The typical redshift of the 10 observed Hot DOGs is $z = 2.7$ \citep{jones14}, higher than the typical redshift of \textit{WISE}/radio AGNs, $z = 1.3$ \citep{jones15}. Although only 10 out of the 30 \textit{WISE}/radio AGN redshifts are spectroscopically known from the SCUBA-2 subset, from \citet{lonsdale15} redshifts for 45 out of 49 \textit{WISE}/radio AGNs are known, and the typical value was $z = 1.7$. 

Follow-up \textit{Spitzer} Infrared Array Camera (IRAC) imaging of a subset of Hot DOGs found an overdensity of galaxies within 1\,arcmin above the number observed in random pointings \citep{assef15}. They also found that Hot DOG environments are as dense as the clusters found by the Clusters Around Radio-loud AGN (CARLA) surveys \citep{wylezalek13,wylezalek14}. 

%However, both populations are found at cosmic times close to the peak of AGN activity and star-formation \citep{genzel00,chapman05,floch05,richards06,hopkins08,magnelli12}, volume density of AGN and ultra-luminous infrared galaxies (ULIRGs)\footnote{Luminous-infrared galaxies (LIRGs), ultra-luminous infrared galaxies (ULIRGs) and hyper-luminous infrared galaxies (HyLIRGs) have characterising total infrared luminosities (8-1000\,$\mu$m) of $L_{8-1000\mu \textrm{m}} > 10^{11}$ L$_\odot$, $L_{8-1000\mu \textrm{m}} > 10^{12}$ L$_\odot$ and $L_{8-1000\mu \textrm{m}} > 10^{13}$ L$_\odot$, respectively \citep{sanders&mirabel96,lonsdale06}} all peak. 

Studying the environments of Hot DOGs and \textit{WISE}/radio AGNs will help to understand the evolution of galaxies and the link with their host galaxy. This paper will explore the clustering and surface number density of the fields to study the environments surrounding the \textit{WISE}-selected AGN. Also the properties of the companion sources around Hot DOGs and \textit{WISE}/radio AGNs will be investigated to determine their nature.  

%see for example \citep{assef15}

%The environments will be compared to each other and also to previous submm surveys. 
%essential to understanding the growth of SMBH during major formation events, and the link with their host galaxy \citep{assef15}. This can help with understanding the processes involved in the evolution of today's galaxies.
%This suggests a link between supermassive black holes (SMBH) growth and star formation rate. 

In Section 2 the surface number density and space density of SMG sources in the fields around Hot DOGs and \textit{WISE}/radio AGNs are compared. In Section 3 the angular two-point correlation function is used to characterise the clustering of the companion SMGs around the \textit{WISE}/radio AGNs.
% along with Monte Carlo simulations.
In Section 4 the properties of the companion sources in the Hot DOG and \textit{WISE}/radio AGN fields are compared using submm, SFR estimations, star formation rate density (SFRD) estimates, mid-infrared (mid-IR) and radio data with previous surveys of companion SMG sources in the Hot DOG and \textit{WISE}/radio-selected AGN fields. The nature of the companion sources detected in the overdense regions of both Hot DOGs and \textit{WISE}/radio AGNs. 
The nature and properties of the companion sources around Hot DOGs and \textit{WISE}/radio AGNs are described in Section 5.

Throughout this paper we assume a $\Lambda$-CDM cosmology with H$_0$ = 71\,km\,s$^{-1}$Mpc$^{-1}$, $\Omega_{\rm{m}}$ = 0.27 and $\Omega_\Lambda$ = 0.73. \textit{WISE} catalogue magnitudes are converted to flux densities using zero-point values on the Vega system of 306.7, 170.7, 29.04 and 8.284\,Jy for WISE 3.4, 4.6, 12 and 22\,$\mu$m wavelengths, respectively \citep{wright10}.

\section{Companion Source Density}

JCMT SCUBA-2 observations of Hot DOGs and \textit{WISE}/radio AGN were in the ``CV DAISY'' mode that produces a uniformly deep coverage 3-arcmin diameter map \citep{holland13}.
Seventeen companion sources were detected at 3\,$\sigma$ significance or above in 10 JCMT SCUBA-2 fields of Hot DOGs reported by \citet{jones14} with an average root mean square (RMS) noise of 1.8\,mJy\,beam$^{-1}$, as shown in Table 1. 

%Each scan was 30 minutes long and four scans were made per Hot DOG target and three scans were made per \textit{WISE}/radio AGN, totalling a exposure time of 120 and 90 minutes per Hot DOG and \textit{WISE}/radio AGN target, respectively. The maps were reduced with STARLINK SubMillimeter User Reduction Facility (SMURF) data reduction package with the Blank Field configuration suitable for low SNR point sources \citep{chapin13}.

\begin{figure}
\begin{centering}
\includegraphics[width=84mm]{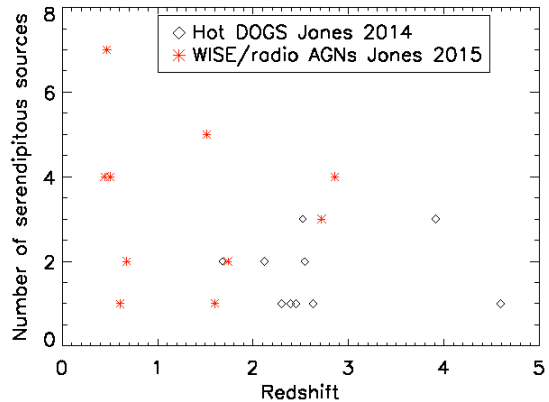}
\caption{The number of submm companion sources found in each of the 10 Hot DOG fields with known redshifts \citep{jones14}, and the 10 \textit{WISE}/radio-selected AGN fields with known redshifts (Jones et al. 2015), are shown by black diamonds and red asteriks, respectively. From previous blank field submm surveys $\sim$ 1 companion source is expected in each 1.5\,arcmin radius SCUBA-2 field.}
\label{redshiftss}
\end{centering}
\end{figure}

Eighty-one companion sources were detected at 3\,$\sigma$ or greater significance in 30 \textit{WISE}/radio-selected AGN fields reported by Jones et al. (2015) with average RMS noise of 2.1\,mJy\,beam$^{-1}$, see Table 2-5. They concluded that they have a higher density of SMGs when compared with Hot DOGs by an additional factor of 2.4 $\pm$ 0.9 \citep{jones15}. The \textit{WISE}/radio AGNs have a lower redshift range, fewer of the \textit{WISE}-selected AGNs are submm detected and lower total IR luminosities compared with Hot DOGs \citep{jones14,jones15}. The lower redshift range and higher overdensity of SMGs around \textit{WISE}/radio AGNs can be seen in Figure~\ref{redshiftss}.  While the observed Hot DOGs have a typically higher redshift than the \textit{WISE}/radio AGNs, the companion sources are matched in submm luminosity (see Tables 1-9) and they are consistent with having similar mid-IR to submm ratios. The K-correction at wavelengths longer than 500microns remains approx. constant with increasing redshift. Due to this K-correction effect the SCUBA-2 fraction of SMG detection should be independent of redshift.  

\begin{figure*}
\begin{centering}
\includegraphics[width=174mm]{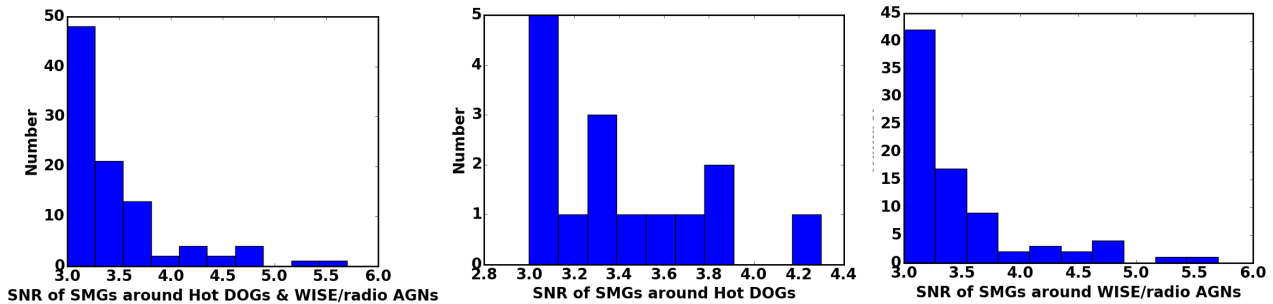}
\caption{Three histograms to show the distribution of SNR of the companion sources detected around Hot DOGs (centre), \textit{WISE}/radio AGN (right) and combining both sets (left).}
\label{snr}
\end{centering}
\end{figure*}

The detection level was set at 3\,$\sigma$ or greater in order to have completeness but reduce the chance of spurious false positive detections. However, there is controversy over whether 3\,$\sigma$ are reliable (e.g. \citep{coppin05,casey12}). Figure~\ref{snr} presents the signal-to-noise ratio (SNR) of the companion sources around Hot DOGs and \textit{WISE}/radio AGN and for the two data sets combined. As expected the higher SNR the fewer sources detected and the less complete the sample. \citet{jones15} looked at the number of SMGs in the \textit{WISE}/radio AGN fields detected at greater than 3\,$\sigma$ and 4\,$\sigma$ and compared to the LABOCA ECDFS Submm Survey (LESS) \citep{weiss09} and concluded the overdensity of SMGs detected above 3\,$\sigma$ is consistent with SMGs detected above 4\,$\sigma$.

%or 3.5\,$\sigma$ significance level is a more robust cut off limit e.g. \citet{coppin05,casey12}. 

Comparing these number counts to ``blank field submm" surveys shows them to be overdense. The blank field submm surveys used to compare were the LESS survey, Cosmological Evolution Survey (COSMOS) \citep{casey13} and the SCUBA Half-Degree Extragalactic Survey (SHADES) \citep{coppin06} fields. The Hot DOG fields have a SMG overdensity by factor of $\sim$ 2 - 3 compared with previous blank field submm surveys , and the \textit{WISE}/radio-selected AGN fields have an even greater overdensity, by a factor $\sim$ 5 - 6 (Jones et al. 2015). However, LESS fields could be under dense by a factor of $\sim$ 2 e.g. \citet{swinbank14}, and the overdensity factor of the Hot DOG fields is less secure, but compared to COSMOS and SHADES there is still an overdensity factor between 2-3.

\begin{figure}
\begin{centering}
\includegraphics[width=84mm]{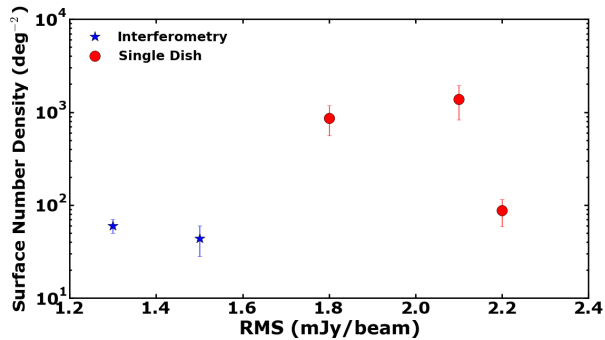}
\caption{The surface number density of SMGs against RMS in the Hot DOG and \textit{WISE}/radio AGN fields are shown in red circles, in comparison to the GOODS-N field single submm dish observations \citep{pope05} also in red circles. The blue stars are submm interferometry observations of SMGs by \citet{toft14} and \citet{smolcic12}. The surface number density of the Hot DOG and \textit{WISE}/radio AGN fields are higher than previous submm surveys.}
\label{rmsover}
\end{centering}
\end{figure}

The surface number density of SMGs in the Hot DOG fields is 866$\pm$210\,deg$^{-2}$, and 1375$\pm$152\,deg$^{-2}$ in the \textit{WISE}/radio AGN fields, to a depth of 1.8\,mJy beam$^{-1}$ and 2.1\,mJy beam$^{-1}$ (submm single dish), respectively. These are higher than previous observations of the surface number density of SMGs, as can be seen in Figure~\ref{rmsover} where SMG surface number densities of different submm surveys are plotted against RMS. \citet{toft14} found at $z \ge 3$ the surface density of bright SMGs is 60$\pm$10\,deg$^{-2}$ to a depth of 1.3\,mJy beam$^{-1}$ (submm interferometry), which was found to be $\sim$30$\%$ lower than $z \sim 2$ quiescent galaxies. SMGs from the LABOCA-COSMOS survey were found to have a surface density between 34$\pm$14\,deg$^{-2}$ and 54$\pm$18\,deg$^{-2}$ at a depth of 1.5\,mJy beam$^{-1}$ (submm interferometry), which was higher than models predicted \citep{smolcic12}. In the GOODS-N field the surface density of SMGs was found to be $\ge 87$\,deg$^{-2}$ \citep{pope05} at depths ranging from 0.3 to 4.1\,mJy beam$^{-1}$ (submm single dish). This was likely higher than previous observations due to the association with a protocluster at $z \sim 4.05$. \citet{geach17} found a tentative overdensity in the GOODS-N compared to the rest of the S2CLS, while combining all of the S2CLS fields, the number counts are consistent with previous studies. Figure~\ref{rmsover} also visually highlights the difference in RMS between single dish and interferometer measurements.

The space density of SMGs in the Hot DOG fields on average is 3.7\,Mpc$^{-3}$ (range 3.0\,Mpc$^{-3}$ to 6.2\,Mpc$^{-3}$) and in the \textit{WISE}/radio AGN fields the average space density is 2.9\,Mpc$^{-3}$ (range 0.7\,Mpc$^{-3}$ to 15\,Mpc$^{-3}$). These space densities are higher than normal star-forming galaxies ($\sim$ 2$\times 10^{-4}$ Mpc$^{-3}$) and local luminous red galaxies ($\sim$ 10$^{-4}$ Mpc$^{-3}$) \citep{wake08}. Previous studies have found SMGs have low number densities of $\sim$ 1-2$\times 10^{-5}$ Mpc$^{-3}$, and are consistent across all redshifts \citep{wilkinson16}. The S2CLS was found to have SMG number densities between 4$\times 10^{-5}$ Mpc$^{-3}$ to 2$\times 10^{-4}$ Mpc$^{-3}$. This confirms that the fields around Hot DOGs and \textit{WISE}/radio AGNs are overdense compared to previous studies of SMGs and normal star-forming galaxies.

\section{Two-point Correlation Function}

The angular two-point correlation function $\omega(\theta)$ is a statistical way to characterise the clustering of galaxies in 2-dimensional (2D) space \citep{connolly98,efstathiou91}. We detect galaxies on a 2D surface and hence we use the angular version of the 3D spatial correlation function \citep{peebles80}. It is the excess probability of finding galaxies separated by $\theta$ as compared with a random distribution. Using the one of the popular estimators described by \citet{landy93};

\vspace{1cm}
$\omega(\theta) = ((|DD| - 2 \times |DR| + |RR|) / |RR|) + \sigma^{2}$
\vspace{0.5cm}

\noindent where $\omega(\theta)$ is the angular correlation function, DD is the number of pairs of galaxies counted in the sample, RR is the number of pairs of galaxies expected in a random distribution, DR is the number of pairs of galaxies counted between the sample and a random distribution and $\sigma$ is the integral constraint \citep{groth77}. The counts have been normalised by dividing by the total number of pairs in each of the three samples; DD, DR, RR.

\vspace{1cm}

\begin{figure}
\begin{centering}
\includegraphics[width=84mm]{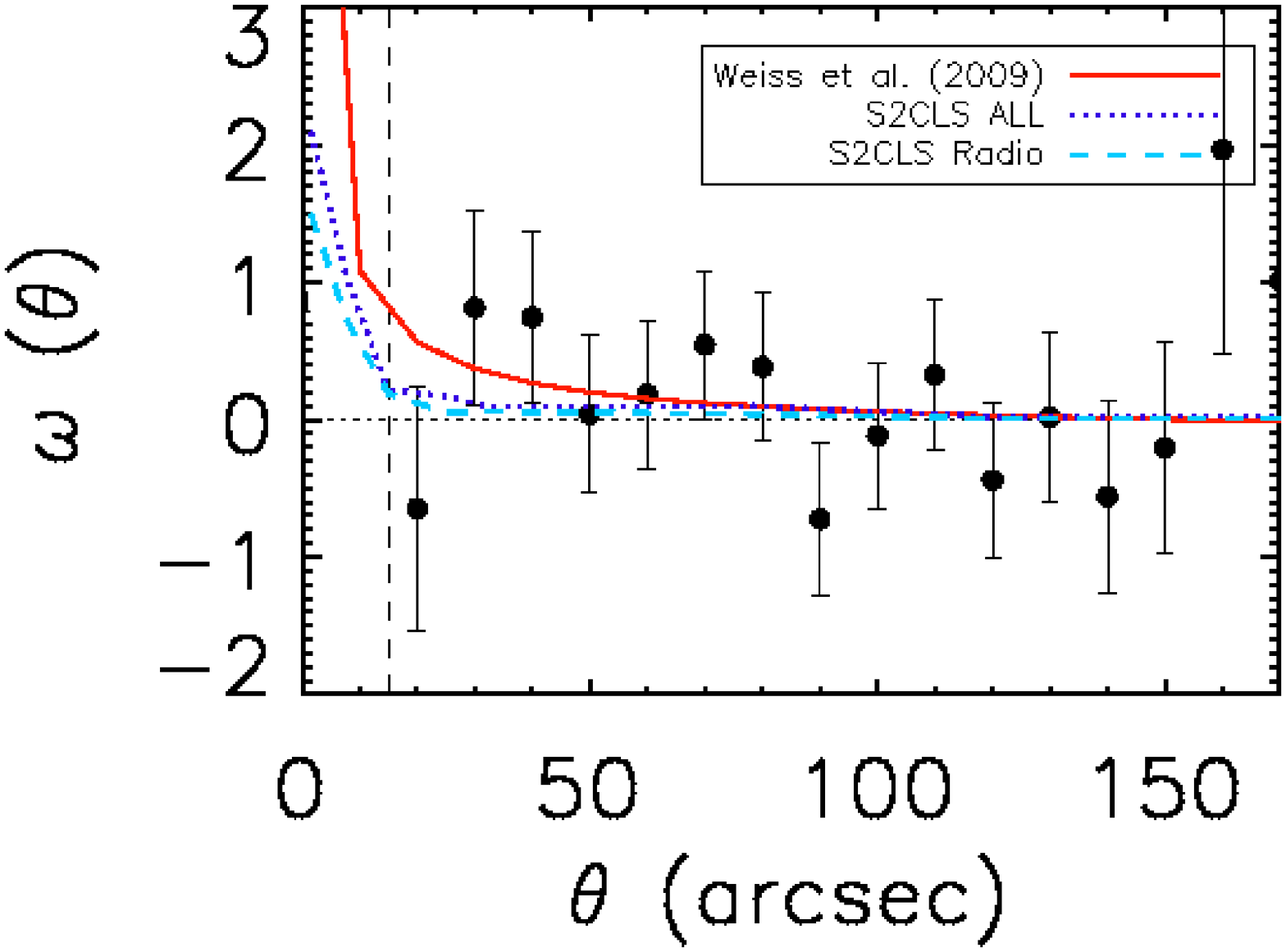}
\caption{Observed angular two-point correlation function using the \citet{landy93} equation. The red solid curve shows the observed angular two-point correlation function for \citet{weiss09}. The blue dotted line and cyan dashed line show the observed angular two-point correlation function for all the SMGs and the subset of radio-detected SMGs, respectively, in the S2CLS \citep{wilkinson16}. Black points represent the observed angular two-point correlation function for the companion sources detected around \textit{WISE}/radio AGNs. The dashed line represents the JCMT SCUBA-2 850\,$\mu$m beam size (15\,arcsec). There were not enough data for reliable results using the Hot DOGs.}
\label{wtheta100000}
\end{centering}
\end{figure}

\noindent The angular correlation function was calculated for the 30 \textit{WISE}/radio-selected AGN fields. It was not calculated for the Hot DOG fields because there were only 10 fields and not a large enough number of companion sources to be statistically significant: the errors would be greater than the large errors on the 30 \textit{WISE}/radio AGNs. To calculate the angular correlation function, 100,000 random fake galaxies were used and compared with the blank-field survey from \citet{weiss09}, that investigated clustering of faint galaxies, see Figure~\ref{wtheta100000}. \citet{weiss09} found significant clustering on scales less than 1\,arcmin and a characteristic angular clustering scale $\theta_0 = 14' \pm 7''$ and a spatial correlation length of $r_0 = 13 \pm 6 h^{-1}$Mpc. We also compared to \citet{wilkinson16} that analysed the largest sample of SMGs (610) in a single field to date from the SCUBA-2 Cosmology Legacy survey (S2CLS) in the redshift range $1 < z < 3$. They found a marginally weaker clustering signal than previous studies, but within 1$\sigma$ uncertainty the results are consistent with \citet{blain04,adelberger05,hickox12}. They also concluded that radio-selected SMGs were slightly more strongly clustered. 

It can be seen in Figure~\ref{wtheta100000} the results in the \textit{WISE}/radio AGN fields provide an upper limit to the strength of an angular clustering signal, and yields a clustering angle of $\theta_0 \geq 80'' $. The clustering signal appears to be inconsistent to previous clustering results of SMGs, however, further observations would provide more definite results. The results from \citet{jones14,jones15} found no evidence for angular clustering when looking at the cumulative fraction of the total number of companion sources in each field within different radii of the \textit{WISE} target and when looking at typical separations compared to Monte Carlo simulations.

%The orange dashed-dotted line shows the angular correlation function of IRAC galaxies selected to match the overlap of the SMGs and galaxies in redshift space \citep{hickox12}. 

\section{Properties of the companion sources around Hot DOGs and \textit{WISE}/radio AGNs}

\subsection{Counterparts of the companion Sources}

A search radius of half SCUBA-2 850\,$\mu$m beam size ($\sim$ 8\,arcsec) was used to find counterparts of these companion SMGs in other catalogues \citep{lilly99,ivison02,hainline09}. This search radius is determined from the probability of finding a source at a given distance from the SMG position \citep{lilly99,ivison02}. This search radius is relatively large due to the difficulty of identifying SMGs at optical and near-IR wavelengths because of the large submm (SCUBA-2) beam, 15\,arcsec at 850\,$\mu$m \citep{dempsey13}. 

Multiple objects within the \textit{WISE} AllWISE Source catalog had two potential counterparts within the 8\,arcsec search radius. To reduce ambiguity in the result the closest in \textit{WISE} W1-W4 bands object is chosen while excluding objects that have \textit{WISE} colours consistent with stars.

\subsection{Mid-IR counterparts}

\begin{figure}
\begin{centering}
\includegraphics[width=84mm]{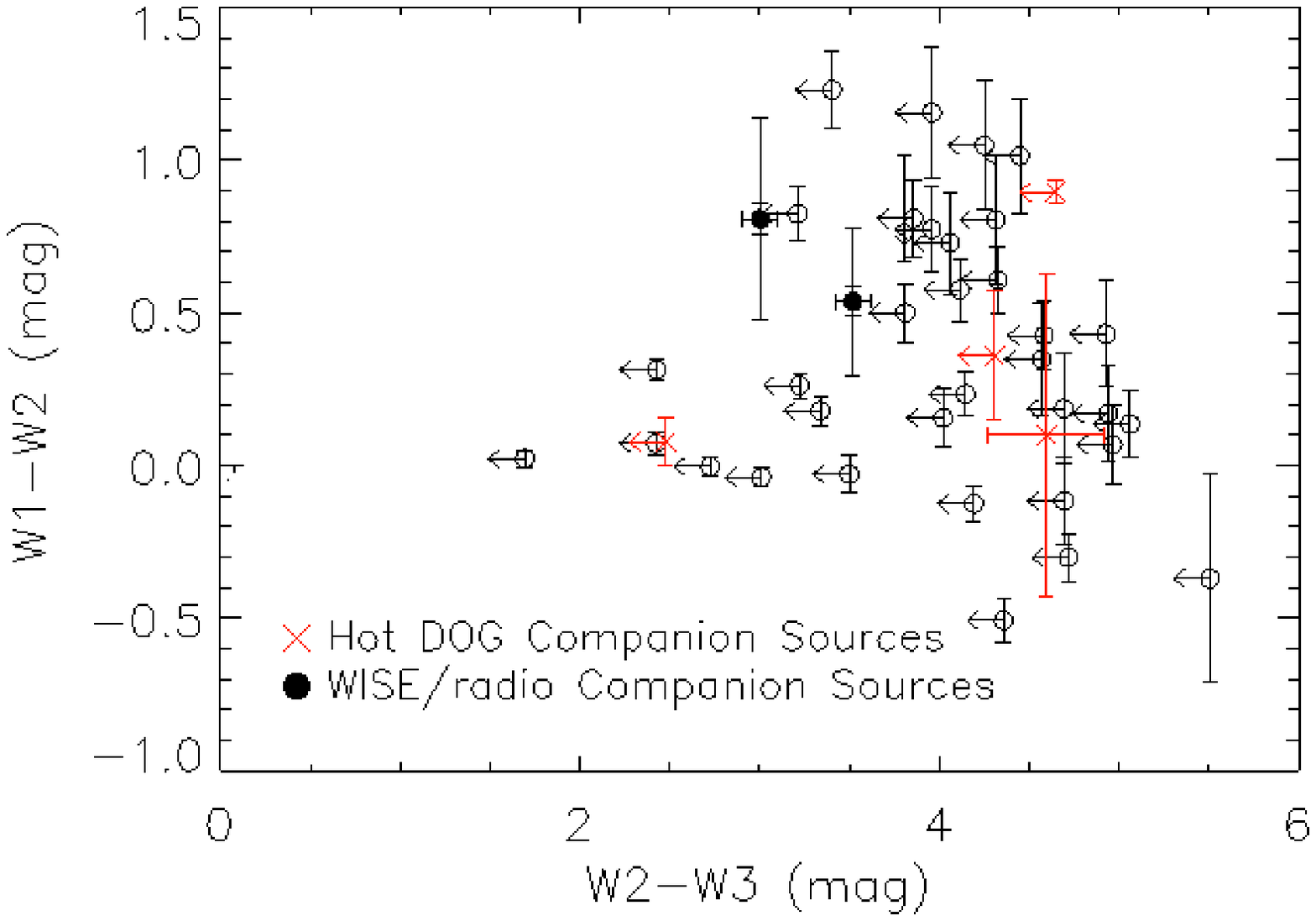}
\caption{\textit{WISE} colour-colour ((4.6\,$\mu$m - 12\,$\mu$m) versus (3.4\,$\mu$m - 4.6\,$\mu$m) or (W2-W3 vs W1-W2)) plot of the Hot DOGs and \textit{WISE}/radio AGNs in red crosses and black circles, respectively. Filled circles are detections. When compared with the \textit{WISE} colour-colour diagram in Figure 12 in \citet{wright10} and Figure 26 in \citet{jarrett11}, the companion sources lie in the starburst zone and appear to be SMGs. They appear to be mid-IR bluer than the Hot DOGs and \textit{WISE}/radio AGNs. }
\label{WISEm13bu02}
\end{centering}
\end{figure}

The \textit{WISE} colour-colour ([W2 - W3] vs [W1 - W2]) plots of the companion sources around Hot DOGs and \textit{WISE}/radio AGNs are shown in Figure~\ref{WISEm13bu02}. These plots can separate different populations of galaxies because of the underlying mechanisms present in each, leading to different mid-IR emission. AGNs are dominated by power-law emission at mid-IR wavelengths. In contrast, normal and star-forming galaxies have a stellar Rayleigh-Jeans tail with additional strong PAH emission, and a continuum that peaks at 70-170\,$\mu$m due to warm dust heated by young stars \citep{jarrett11}. 

Both sets of companion sources have similar \textit{WISE} colours. However, most have upper limits in the W3 band and so have limits to their red W2 - W3 colour. When comparing with the \textit{WISE} colour-colour diagram of different galaxy populations in Figure 12 in \citet{wright10} and Figure 26 \citet{jarrett11}, the companion sources lie in both the starburst (star-forming) galaxy zone and AGN zone. 

\begin{figure}
\begin{centering}
\includegraphics[width=84mm]{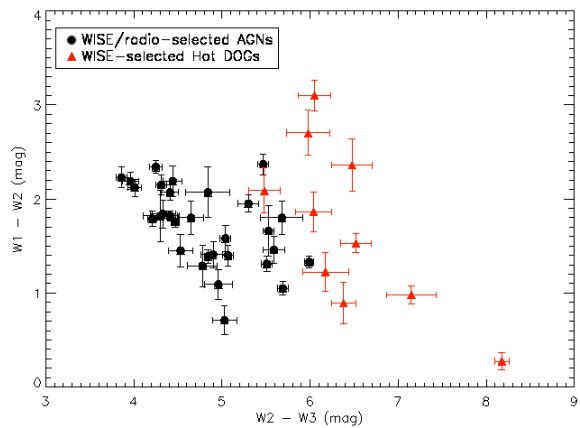}
\caption{\textit{WISE} colour-colour ((4.6\,$\mu$m - 12\,$\mu$m) versus (3.4\,$\mu$m - 4.6\,$\mu$m) or (W2-W3 vs W1-W2)) plot of the Hot DOGs and \textit{WISE}/radio AGNs companion sources in red triangles and black cricles, respectively. When compared with the \textit{WISE} colour-colour diagram in Figure 12 in \citet{wright10}and Figure 26 in \citet{jarrett11}, the \textit{WISE}-selected AGNs are extremely red compared to other galaxy populations. }
\label{wiseradiow1w2}
\end{centering}
\end{figure}

The Hot DOGs and \textit{WISE}/radio AGNs are redder than the companion sources, see Figure~\ref{wiseradiow1w2}. This is no surprise because they were selected to be red \citep{eisenhardt12,lonsdale15}, which could imply they have higher dust obscuration and/or a higher AGN contribution, and higher dust temperatures than that of their companion sources . Hot DOGs and \textit{WISE}/radio AGNs are predominantly powered by AGN \citep{wu12,jones14,jones15,tsai15,lonsdale15}. SMGs are predominantly powered by star formation \citep{alexander05}, and have SEDs dominated by cooler dust emission (20 - 50\,K) \citep{hainline09}.

\citet{hainline09} observed 73 radio-selected SMGs with known redshifts using \textit{Spitzer} IRAC and MIPS and detected 91\,$\%$ at 3.6\,$\mu$m, 91\,$\%$ at 4.5\,$\mu$m, 78\,$\%$ at 5.8\,$\mu$m, 74\,$\%$ at 8\,$\mu$m, 71\,$\%$ at 24\,$\mu$m, and 7\,$\%$ at 70\,$\mu$m. They found that the detection rate in the shortest bands is less than SMGs in deeper \textit{Spitzer} Wide Area Infrared Extragalactic Legacy Survey (SWIRE).  These are higher mid-IR detection rates than the companion SMG sources in the All\textit{WISE} Source Catalog presented in this paper \citep{jones15}; where 24$\%$ and $35\%$ companion SMG sources in the Hot DOG and \textit{WISE}/radio AGN fields, respectively, were detected. The difference in detection rates could be due to the depth of coverage. \textit{Spitzer} IRAC had deeper mid-IR observations with sensitivity ranging from 0.1 - 0.9\,$\mu$Jy at 3.6$\mu$m and 0.4 - 1.8\,$\mu$Jy at 4.5$\mu$m, which are deeper than All\textit{WISE} Source Catalog 0.08\,mJy at 3.4$\mu$m and 0.11\,mJy at 4.6$\mu$m at 2\,$\sigma$. 

\subsection{Submillimetre Emission, SFRs and SFRDs}

The average submm flux density of SMGs around Hot DOGs is S$_{850 \mu m}$ $ =  $ 6.2 $\pm$ 1.8 mJy, which is comparable to SMGs around \textit{WISE}/radio AGNs, S$_{850 \mu m}$ = 7.2 $\pm$ 2.1 mJy. Submm flux densities provide a reliable measurement of SFR \citep{alexander16}. Submm flux densities can be converted to SFRs for SMGs with $z > 1.5$ using

\vspace{1cm}
SFR$_{850 \mu m}$ = 200 $\times$ S$_{850 \mu m}$
\vspace{0.5cm}

\noindent \citep{barger14}. The average SFR is $\simeq$ 1240\,M$_{\odot}$yr$^{-1}$ for SMGs around \textit{WISE}/radio AGNs, slightly lower than the SFR $\simeq$ 1460\,M$_{\odot}$yr$^{-1}$ for SMGs around Hot DOGs. %\citet{barger14} found SMGs with high radio powers, $\ge$ 5 $\times$ 10$^{30}$ erg s$^{-1}$ Hz$^{-1}$, were distinct from other radio sources. However, the SMGs in both \textit{WISE}/radio AGNs and Hot DOGs have similar submm and radio flux densities, see Tables 1-10, and hence it is not surprising they have similar SFRs.

Cosmological simulations predict that overdense regions, $\sim$ 5$\sigma$ density peak, are associated with high SFRs, $\sim$ 750\,M$_{\odot}$yr$^{-1}$ \citep{yajima15}. Observations at redshifts $z \sim 1$ found higher SFRs are associated with higher densities \citep{cooper07}. The mean SFR at the core of protoclusters have been found to be enhanced, up to a factor $\sim$ 5.9 over the field \citep{alexander16}, and outside of the central region the SFR is consistent with field galaxies. ALMA observations of the SSA22 protocluster at redshift $z = 3.09$ found enhanced SFR in the densest regions \citep{umehata15}. Therefore, higher SFRs of the SMGs around \textit{WISE}/radio AGNs than the SMGs around Hot DOGs is expected.
%However, \citep{alexander16} found that the SFR is significantly enhanced by a single extreme object that could be the progenitor of the brightest cluster galaxy with SFR of $\sim$ 1600\,M$_{\odot}$yr$^{-1}$. 

The star formation rate density (SFRD) represents the total star formation transpiring per unit time and volume at a given redshift, as seen in Figure~\ref{sfrd}. SFRD allows direct comparison of the importance of IR-luminous galaxies to the build-up of stellar mass in the Universe. From previous work the SFRD in clusters increases with redshift from $ z \sim 1 $ to $ z \sim 3$ e.g. \citep{hopkins06,bouwens11,magnelli11,clements14}, e.g. \citet{dannerbauer14} measured an SFRD of $\sim 900$ M$_{\odot}$ yr$^{-1}$ Mpc$^{-3}$ in the field around the spiderweb radio galaxy at redshift $z = 2.16$ in a region of 2\,Mpc. However, there are observations of high-redshift clusters with a combination of quiescent and star-forming galaxies \citep{gobat13,strazzullo13}, and clusters dominated by quiescent galaxies \citep{tanaka13}. Therefore, higher redshift SFRDs are needed to understand the history of galaxy clusters especially in the peak epoch of star formation at redshifts $1 < z < 3$, which includes this paper. The SFRDs were calculated for each cluster, which is the \textit{WISE}-selected source and its surrounding SMGs, assuming the SMGs are at the same redshift as the source. We assumed that each cluster was spherical, and derived an angular radius from the SCUBA-2 map, 1.5\,arcmin. The angular radius was converted to a proper distance at the redshift for each cluster, where the redshift is unknown the average redshift is assumed, $z = 1.7$ for \textit{WISE}/radio AGNs. The volume for each cluster was calculated by assuming this proper distance is the radius of the cluster. The SFRDs are presented in Table 6, these are lower limits because faint SMGs could be missed due to the shallow depths of the SCUBA-2 maps.

The SFRDs range for Hot DOGs from 1523$\pm$30M$_{\odot}$ yr$^{-1}$ Mpc$^{-3}$ to 7949$\pm$159 M$_{\odot}$ yr$^{-1}$ Mpc$^{-3}$, and average 3533 M$_{\odot}$ yr$^{-1}$ Mpc$^{-3}$. These are lower than \textit{WISE}/radio AGNs with a range from 1219$\pm$49 M$_{\odot}$ yr$^{-1}$ Mpc$^{-3}$ to 18715$\pm$374  M$_{\odot}$ yr$^{-1}$ Mpc$^{-3}$, and average 3929 M$_{\odot}$ yr$^{-1}$ Mpc$^{-3}$. Our results can be compared to Figure 15 from \citet{clements14} and the SFRDs calculated in this paper are higher than field galaxies from \citet{hopkins06} and \citet{bouwens11}, with SFRDs of $\sim 900$M$_{\odot}$ yr$^{-1}$ Mpc$^{-3}$ and $\sim700$M$_{\odot}$ yr$^{-1}$ Mpc$^{-3}$ at $z = 2$ respectively. Our values are similar to four \textit{Herschel} Multitiered Extragalactic Survey (HerMES) clusters of dusty, star-forming galaxies at redshifts between $z = 0.76$ to $z = 2.26$, and other clusters with MIR/FIR measurements from the literature with SFRDs ranging from $\sim$200M$_{\odot}$ yr$^{-1}$ Mpc$^{-3}$ to $\sim$3000M$_{\odot}$ yr$^{-1}$ Mpc$^{-3}$. Simulations of massive galaxy clusters cannot account for the overdensity found by \citet{clements14}, which is thought to be due to insufficient peaks of star formation activity in the simulations at early epochs, and including strong starbursts in the simulations is required to explain the statistical properties of SMGs \citep{granato15}.

Dusty star forming galaxies (DSFG)-rich protoclusters at redshifts $2 < z < 3$ were shown to have slightly higher SFRDs compared to the field, due to their large occupying volumes \citep{casey16}. In contrast virialised clusters at redshifts $z < 1$ have a substantially higher SFRD. This is in agreement with the lower redshift \textit{WISE}-selected AGNs W0342$+$3753, W1501$+$1324 and W2230$-$0720 that have a redshift of $ z = 0.47$, $z = 0.505$ and $ z = 0.444$, respectively and a significantly higher SFRD at 18715$\pm$374 M$_{\odot}$ yr$^{-1}$ Mpc$^{-3}$, 9501$\pm$190 M$_{\odot}$ yr$^{-1}$ Mpc$^{-3}$ and and 12596$\pm$252 M$_{\odot}$ yr$^{-1}$ Mpc$^{-3}$, respectively. This is due to a high overdensity of SMGs in each field; seven, four and four serendipitous SMG sources in each field respectively.

Completeness is the rate at which a source is expected to be detected in a map \citep{hatsukade13}. It is computed by simulating the detection rate of 1,000 fake point sources per flux bin placed in the real cleaned signal map \citep{tamura09}. Brighter SMGs where the flux density is S$_{1100\mu m}$ $\ge 2.7$\,mJy were found not to be significantly affected from incompleteness and false detections \citep{tamura09}. They found that the completeness was $\sim$ 50\,$\%$ at 2.7\,mJy and 90\,$\%$ at 4.0\,mJy. All the companion sources detected around Hot DOGs and \textit{WISE}/radio AGNs have flux densities S$_{850\mu m}$ $\ge 4.6$\,mJy and S$_{850\mu m}$ $\ge 5.5$\,mJy, respectively see Tables 1-5. The completeness was found to range from 77\,$\%$ and 100\,$\%$ reported by \citet{hatsukade13} from 15 SMGs observed. It was concluded that the correction for incompleteness and contamination has an effect on the low flux density bins (S$_{850\mu m}$ $< 2.9$\,mJy) and a minimal effect on the high flux density bins S$_{850\mu m}$ $\ge 2.9$\,mJy \citep{casey13}. This is also confirmed by \citet{weiss09} where the source extraction is complete ($>$ 95\,$\%$) for sources with flux densities S$_{870\mu m}$ $\ge 6.5$\,mJy and 50\,$\%$ complete at $\sim$ 4.0\,mJy. Therefore, the SMG completeness of the fields around Hot DOGs and \textit{WISE}/radio AGNs should be between 50 and 100\,$\%$ complete. Hence the SFRDs could be higher than calculated here. This will also have an effect on the number count comparison with other submm surveys, where the overdensities in the \textit{WISE}-selected Hot DOG and \textit{WISE}/radio-selected AGN fields could have an even higher overdensity.

%The SFRDs in the 2\,Mpc region range from 74.0 M$_{\odot}$ yr$^{-1}$ Mpc$^{-3}$ to 248.4 M$_{\odot}$ yr$^{-1}$ Mpc$^{-3}$ for Hot DOGs, and 57.3\,Mpc to
\begin{figure}
\begin{centering}
\includegraphics[width=84mm]{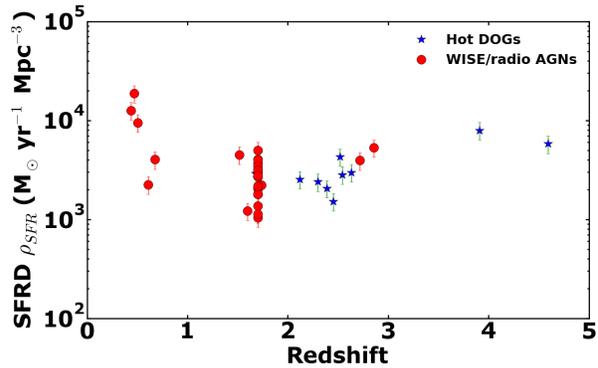}
\caption{Star formation rate density (SFRD) of each Hot DOGs and \textit{WISE}/radio AGNs and their surrounding SMGs. The SFRDs are higher than field galaxies and similar to the four HerMES clumps, other clusters with MIR/FIR measurements from the literature, when comparing with Figure 15 from \citet{clements14}. }
\label{sfrd}
\end{centering}
\end{figure}

%None of the companion SMG sources in the Hot DOG and \textit{WISE}/radio AGN fields were detected in the NVSS and/or FIRST radio catalogues.

%or in \textit{Herschel} images of the Hot DOGs catalogue (priv. comm. J. Wu). 

\subsection{Radio Emission}

None of the companion sources around Hot DOGs or \textit{WISE}/radio AGNs were detected at radio wavelengths in FIRST and/or NVSS, where the typical 1.4\,GHz detection limit was 1.0\,mJy/beam, see Table 1-5. From previous observations $\sim$ 65\,$\%$ of SMGs with submm flux densities S$_{850\mu m}$ $>$ 5\,mJy had detectable radio emission in much deeper observations with RMS $\sim$ 10\,$\mu$Jy (range 2.3 to 17.4\,$\mu$Jy) \citep{ivison98,ivison02,chapman05}. 
%(range 2.3 to 17.4\,mJy) 

None of the 17 or 81 companion sources in the Hot DOG and \textit{WISE}/radio AGN fields, respectively, are detected in NRAO snapshot follow-up VLA radio maps (priv. comm. with C. Lonsdale). The non-detections are consistent with SMG SEDs at relevant redshifts.
%No radio data could suggest cooler dust temperatures than radio-detected SMGs \citep{hainline09}. This is not surprising because the NVSS survey is a set of 2326 continuum fields each covering 4\,deg$^2$ and contains $\sim$ 1.8 million discrete sources. This would imply a source density of 48 sources\,deg$^{-2}$, and therefore we would expect $\sim$0.9 and $\sim$ 2.8 NVSS sources in the 10 Hot DOGs and 30 \textit{WISE}/radio AGN SCUBA-2 fields, respectively. 

\subsection{X-ray Emission}

No counterparts to the companion sources from point sources were found in the third \textit{XMM-Newton} companion Source Catalog, \textit{3XMM}-DR5 \citep{rosen15}. However, previous deeper X-ray observations of SMGs found only 16 with X-ray detections from sample size of 35 (45$\pm$8\%) \citep{laird09}. These observations were from the 2-Ms \textit{Chandra} survey with flux limits on the order of 10$^{-17}$erg cm$^{-2}$ s$^{-1}$ and a range 1.1 to 17.7$\times$10$^{-16}$erg cm$^{-2}$ s$^{-1}$ , and are deeper compared with \textit{3XMM}-DR5 of the order 10$^{-15}$erg cm$^{-2}$ s$^{-1}$ for 3 sigma detections. Therefore, non-detections from point sources are expected.

\section{Discussion}

\subsection{Companion Source Clustering}

Figure~\ref{wtheta100000} provides an upper limit to the strength of an angular clustering signal in the \textit{WISE}/radio AGN fields, and appears to be inconsistent with previous clustering studies of SMGs from \citet{weiss09,hickox12,wilkinson16}. \citet{weiss09} found consistent correlation length values of SMGs with \citet{blain04,farrah06} but inconsistent with \citet{scott06}, this could be explained by the small significance of the clustering signal in both studies. 

\citet{hickox12} reanalysed SMGs from the LABOCA survey in a novel method to cross-correlate SMGs in the LABOCA survey and galaxies from \textit{Spitzer} IRAC. They found a lower correlation length, $r_0 =  7.7^{+1.8}_{-2.3}$\,$h^{-1}$Mpc at $z = 2$, than \citet{weiss09}, but one that is consistent with measurements for optically-selected QSOs. The observed clustering could depend on the submm flux limit of the survey, presence of redshift spikes and uncertainties in redshift selection function \citep{williams11,adelberger05}, which could result in uncertainties in clustering estimates. \citet{hickox12} compared their autocorrelation length $r_0$ to previous SMG results with a range of 850$\mu$m flux limit from 3 to 6\,mJy, and found consistent angular clustering estimates. They concluded that SMGs are likely to represent a short-lived transition phase from cold, gas-rich, star-forming galaxies to passively evolving systems.

\citet{wilkinson16} found when analysing the largest sample of SMGs in the S2CLS, SMGs are not as strongly clustered as previously thought. However, their measurements were in agreement with previous studies \citet{blain04,hickox12} within 1$\sigma$ errors, and found a weaker clustering signal when comparing to \citet{weiss09}. Accounting for blending could bring the previous studies into better agreement with \citet{wilkinson16}.  Alternatively, the SMG clustering could depend on redshift, large-scale environment and merger history. They found that the clustering of SMGs are consistent with star-forming population and lower than passive population at the same redshift, and tentative evidence of halo downsizing. \citet{chapman09} proposed that SMGs do not necessary trace the most massive dark matter halos. 

\citet{donoso14} analysed the angular clustering properties of a sample of $\sim$ 170,000 \textit{WISE}-selected AGN with very red mid-IR colours. The whole sample were found to have a similar clustering strength to optically-selected quasars at comparable redshifts ($z = 1.1$) in the Sloan Digital Sky Survey (SDSS) \citep{porciani04,croom05,myers07}. They are found in denser environments when compared with all SDSS galaxies at that redshift. Redder AGN which are well detected at 4.6\,$\mu$m (W2) have a stronger clustering bias (relationship between the distribution of dark matter and luminous matter) than blue AGN. \textit{WISE}/SDSS obscured AGN are more strongly clustered and inhabit denser environments than unobscured AGN. \citet{dipompeo2014} confirmed this but found a smaller difference in angular clustering amplitude between \textit{WISE}-selected obscured quasars and unobscured quasars. However, \citet{mendez15} found no significant difference between obscured and unobscured AGN.

There is an overdensity of SMGs with $\sim$ 2 or 3 SMGs per SCUBA-2 field compared with the expectation of 1 SMG from blank field submm surveys. The number of sources for the angular two-point correlation function of Hot DOGs and \textit{WISE}/radio AGN fields were not numerous enough to see an angular clustering clustering signal. Monte Carlo simulations of the typical separation of the companion sources and the cumulative fraction of the total number of companion sources within different radii from the \textit{WISE} target showed no angular clustering. This is agreement with \citet{assef15} that found no angular dependence of the IRAC overdensities around a subset of Hot DOGs.

From previous evidence there could be clustering on scales greater than the SCUBA-2 fields \citep{scoville00,blain04,greve04,farrah06,ivison07,weiss09,cooray10,scott10,hickox12}. Alternatively, the clustering peak could be off centre from the \textit{WISE} source and not on the SCUBA-2 1.5\,arcmin map scale. This agrees with \citet{smail14} where overdensities of the most active, ultraluminous star-forming galaxies were offset from the assumed protocluster centre and are situated in the lower-density environments. \citet{dannerbauer14} observed a density of SMGs up to four times greater than in blank field surveys that were not centred on the submm-bright radio galaxy.

 \citet{muldrew15} explored the structures of protoclusters and their relationship with high-redshift galaxies using the Millennium Simulation. They found that protocluster structures are very extended at the redshifts ($z = 2$) we are probing with 90$\%$ of their mass is dispersed across $\sim$ 30\,arcmin ($\sim$ 35\,$h^{-1}$ Mpc comoving). This would imply that many observations of protoclusters and high-redshift clusters are not imaging all of the cluster. Many protoclusters have no central or main halo that could be classified as a high-redshift cluster, only 10$\%$ were dominated by a single halo at redshift $z = 2$. This could explain why there is no evidence or only an upper limit of angular clustering in the Hot DOGs and \textit{WISE}/radio AGNs fields on $\sim$ 1.5\,arcmin scales from Monte Carlo simulations of typical separations. Alternatively, the cluster might be peaked substantially off-centre from the \textit{WISE} target. Further observations of companion sources in the fields around \textit{WISE}/radio AGN are needed to determine the angular two-point correlation function $\omega(\theta)$. Wide-field sub(mm) surveys are needed to cover the total (proto)cluster structure, and is in agreement with results from \citet{casey16,hung16}, where \citet{casey16} found protoclusters subtend 10\,arcsec to a half degree in the sky and at redshifts $z \ge 2$ their overdensity is difficult to detect due to their large occupied volumes. \citet{hung16} found large scale structure around a cluster to within 10\,arcsec.

%with no peaking up on the \textit{WISE} source on arcmin scales
\citet{viero13} presented observations from \textit{Herschel} and found a clustering signature from SMGs that could be decomposed into 2-halo (linear) power from galaxies in separate halos, and 1-halo (non-linear) power from multiple central and satellite galaxies occupying massive halos. It has been found that a fraction of luminous sources are found within these satellite halos for example \citet{gonzalez11} predicts 38$\%$ SMGs and 24$\%$ SMGs with S$_{850 \rm{\mu m}} > 5$\,mJy are satellites. Additionally, star-forming galaxies in groups and clusters were found in the outskirts of massive cluster-scale halos \citep{muldrew15}. The lack of clustering signal of SMGs in the Hot DOG and \textit{WISE}/radio AGN fields could be because they are also in the outskirts of diffuse massive halo and not having enough sources.

\subsection{Companion Source Properties}

%however, for the detected sources their \textit{WISE} colour properties are consistent with SMGs, and not AGN dominated. 

Only a fraction of the SCUBA-2 companion sources are detected in \textit{WISE} . The \textit{WISE} colours of the companion sources are consistent with star-forming galaxies and AGN, while their mid-IR to submm ratios are not consistent with AGN dominated sources \citep{jones15}. The companion sources hence appear to be consistent with SMGs. The SMG SFRs were estimated using their submm flux densities and are consistent with SMGs; the average SFR is $\sim$ 1240\,M$_{\odot}$yr$^{-1}$ for SMGs around \textit{WISE}/radio AGNs, slightly lower than the SFR $\sim$ 1460\,M$_{\odot}$yr$^{-1}$ for SMGs around Hot DOGs. The SMGs around \textit{WISE}/radio AGNs have slightly higher SFRs than around Hot DOGs by $\sim$ 18$\%$, which is expected that SFRs are enhanced in denser regions.

When comparing the companion SMG sources radio properties to previous SMGs, around 65 - 70\,$\%$ of bright SMGs (S$_{850\mu m}$ $>$ 7\,mJy) have been detected at S$_{1.4\rm{GHz}}$ \citep{ivison02,borys04}. It has been suggested that the radio-undetected SMGs may have colder dust or lie at $z > 3$ \citep{ivison02,eales03,swinbank08}. No companion sources have radio detections in shallow NVSS or FIRST images, and the radio data are not deep enough to assess their dust temperatures.

%when comparing the cumulative number counts and typical separations of the companion SMG sources to Monte Carlo simulations, or comparing the angular correlation function to previous SMG \citep{weiss09} and extremely red object \citep{daddi07} surveys

%Hot DOGs and \textit{WISE}/radio AGNs are consistent with very luminous, AGN-dominated galaxies. They both appear to be in overdense extended regions of SMGs dominated by starburst activity. The companion SMG sources are mid-IR bluer in \textit{WISE} colours and hence have cooler dust temperatures than Hot DOGs and \textit{WISE}/radio AGNs (60 - 120\,K; Wu et al. 2012; Bridge et al. 2013; Lonsdale et al. 2015).

% \textit{WISE}/radio AGNs could be the best signposts of overdense regions in the sky. 

The SFRDs of the \textit{WISE}-selected AGNs are higher than the field but consistent with measurements of clusters of dusty galaxies from HerMES and DSFGs or luminous AGN. Conclusions from observations of $z > 2$ protoclusters suggest that the universe's largest galaxy clusters are thought to built by massive $> 10^{11}$ M$_{\odot}$ galaxies in short-lived bursts of activity. The challenge has been to observe these structures when they have such large volumes, subtending $\sim$ 0.5 degrees on the sky \citep{casey16}. However, the \textit{WISE}-selected AGNs have high SFRDs with consistent values to these previous observations of clusters of dusty star-forming galaxies, but are on smaller volumes, with a SCUBA-2 map radius of 1.5\,arcmin. Therefore, \textit{WISE}-selected AGNs could be used to study protoclusters at high redshift on small volumes (arcmin scales) of the sky.

%Further observations are needed to increase the sample size to be able to draw more accurate conclusions of whether they represent a different phase of galaxies in the major merger model, or if they are very luminous class of DOGs. As well as more spectroscopic redshifts of the targets and their companion sources, to see if the companion sources are at the same redshift as the targets. Therefore, whether the \textit{WISE}-selected targets are in overdense regions of the sky. \textit{Herschel} observations of \textit{WISE}-selected AGNs near the peak of their SED will be presented in further papers (Tsai et al. 2015). High-resolution ALMA data are needed to resolve the galaxies to see if there are multiple components in \textit{WISE}-selected targets. 
%For example ALMA data of \textit{WISE}-selected Hot DOG W2026$+$0716 is needed because it was found to have an increase in flux density in the 850\,$\mu$m SCUBA-2 map up to diameter of 29\,arcsec and could be a sign of multiple components \citep{jones14}.

\section{Summary}

Previously Hot DOGs and \textit{WISE}-selected AGNs were found to be extremely obscured, hyperluminous AGN at redshifts between $0.4 < z < 4.6$. Their environments were found to be overdense in SMGs and these overdensities have been investigated here.

% The space densities, SFR, SFRDs and angular clustering signal of the SMGs have been used to compare to previous dusty galaxy populations  and their overdensities. 

\begin{itemize}

%\item{Hot DOGs and \textit{WISE}/radio AGNs have very high total IR luminosities, hot dust temperatures (60 - 120\,K), and SEDs that are not well fitted by many standard AGN templates due to excess mid-IR emission and less submm emission.}

%\item{Hot DOGs and \textit{WISE}/radio AGNs appear to be consistent with the same population of very luminous, AGN-dominated galaxies but are different redshifts. They could be a new transient phase of the major merger model.}

%\item{The companion sources detected around Hot DOGs and \textit{WISE}/radio AGNs have \textit{WISE} colours consistent with star-forming galaxies and mid-IR to submm ratios not consistent with AGN dominated sources. This could imply that they are all consistent with SMGs.}

%, \textit{WISE} W3 (12\,$\mu$m) versus \textit{WISE} (W1 - W2) (4.6\,$\mu$m - 12\,$\mu$m) and W4 (22\,$\mu$m) flux density versus SCUBA-2 850\,$\mu$m flux density,

%\item{Their properties including luminosity, SFR, stellar mass, are consistent with previous observations of SMGs}
% have lower detection rates in the \textit{WISE} bands when compared to previous mid-IR studies of SMGs using IRAC. This could be due to shallower mid-IR observations of \textit{WISE}. Their position of the companion sources on the

%\item{All the companion sources have bluer mid-IR positions in the \textit{WISE} colour-colour plot compared with Hot DOGs and \textit{WISE}/radio AGNs, which implies cooler dust temperatures than 60 - 120\,K.}

\item{The space densities of SMGs around the \textit{WISE}-selected AGNs were found to overdense compared to normal star-forming galaxies and SMGs in the S2CLS.}

\item{The SMGs around \textit{WISE}/radio AGNs $\sim$ 18$\%$ higher SFRs than SMGs around Hot DOGs.}

\item{The SFRDs of the \textit{WISE}-selected AGNs are higher than field galaxies, and consistent with values for known clusters of dusty galaxies.}

%\item{There was no radio emission data at mJy limit for the companion sources and implies cooler dust temperatures than radio-detected SMGs with T $\sim$ 20 - 30\,K.}

%\item{\textit{WISE}/radio AGNs are typically at a lower redshift ($z = 1.7$) than Hot DOGs ($z = 2.7$). The lower redshift \textit{WISE}/radio AGNs appear to reside in higher density regions compared with higher redshift Hot DOGs. This could be due to differences in redshift and/or radio emission. However, more observations are needed because only 10 targets in each sample have known redshifts.}

%\item{The higher radio emission of \textit{WISE}/radio AGNs could indicate that AGN with powerful radio jets reside in more overdense regions than radio-quieter AGNs, that could be due to major mergers producing rapidly spinning SMBHs, or the surrounding IGM enhancing synchrotron emission from radio jets.}

%\item{Alternatively the higher overdensity of \textit{WISE}/radio AGNs compared to Hot DOGs could be due to their lower typical redshift at $z = 1.7$ and is closer the peak redshift distribution of SMGs found from previous surveys of $z = 2.2 \pm 0.1$ \citep{chapman05,clements08,dey08,wardlow11}.}

\item{The results impose an upper limit to the strength of angular clustering of the companion SMG sources in Hot DOGs and \textit{WISE}/radio AGNs on SCUBA-2 1.5\,arcmin scales. The typical separations when compared to Monte Carlo simulations showed no angular clustering. This is an agreement with the cumulative fraction of companion sources in different radii from the \textit{WISE} target. This could be because they are satellite galaxies in the massive halo or that the protocluster is on bigger scales (up to $\sim$ 30\,arcmin) and we are not fully probing the protocluster.}

\item{Hot DOGs and \textit{WISE}/radio AGNs appear to be signposts of overdense environments.}

%The \textit{WISE}-selected AGN physical properties are unusual compared to field galaxies and may represent an extreme transitional phase in the galaxy evolution model.
%\item{Further spectroscopic redshift data of the \textit{WISE}-selected targets and their companion SMG sources are needed.}

%\item{Further submm data of \textit{WISE}-selected targets are needed to increase the sample size of \textit{WISE}-selected targets. Also high-resolution ALMA data are needed to resolve the galaxies to see if there are multiple components for example of \textit{WISE}-selected Hot DOG W2026$+$0716.}

\end{itemize}

\section{Acknowledgements}

The authors thank the anonymous referee for his/her comments and suggestions, which have greatly improved this paper.
S.F.J. thanks Kirsten K. Knudsen for inspiring comments and helpful advice.

This publication makes use of data products from the \textit{Wide-field Infrared Survey Explorer}, which is a joint project of the University of California, Los Angeles, and the Jet Propulsion Laboratory/California Institute of Technology, funded by the National Aeronautics and Space Administration. The James Clerk Maxwell Telescope has historically been operated by the Joint Astronomy Centre on behalf of the Science and Technology Facilities Council of the United Kingdom, the National Research Council of Canada and the Netherlands Organisation for Scientific Research. Additional funds for the construction of SCUBA-2 were provided by the Canada Foundation for Innovation. The program IDs under which the data were obtained were M12AU10, M12BU07 and M13BU02. RJA was supported by FONDECYT grant number 1151408. This work is based in part on observations made with the \textit{Spitzer Space Telescope}, which is operated by the Jet Propulsion Laboratory, California Institute of Technology under a contract with NASA

\bibliographystyle{mn2e}
\bibliography{ref}

\setlength{\tabcolsep}{4pt}

\begin{landscape}
\begin{table*}
\caption[Coordinates and photometry of the 17 companion SMG sources found around 10 Hot DOGs.]{Coordinates and photometry of the 17 companion SMG sources found around 10 Hot DOGs, with 3.4\,$\mu$m, 4.6\,$\mu$m, 12\,$\mu$m and 22\,$\mu$m magnitudes from the All\textit{WISE} Source Catalog and 850\,$\mu$m flux densities from SCUBA-2. The targets with \textit{WISE} upper limits have SNR $<$ 2 and therefore in the All\textit{WISE} Source Catalog the magnitudes quoted are 2\,$\sigma$ upper limits. The top six \textit{WISE}-selected HotDOGs are detected at 850\,$\mu$m, while the bottom four Hot DOGs have upper limits at 850\,$\mu$m. No \textit{Herschel}, NVSS, FIRST data and no objects found in SIMBAD for all the 17 companion SMG sources. FIRST detection limit for each source position is given in mJy/beam, "undetected" represents that there is no FIRST coverage or NVSS detection. A search radius of 8\,arcsec was used, which is a typical search radius of SMG counterparts \citep{hainline09}. $^a$Subset of these Hot DOGs have been reported by \citet{tsai15}. $^b$Subset of these Hot DOGs have been reported by \citet{eisenhardt12}.}
\hspace*{-4cm}\begin{tabular}{@{}cccccccccccc@{}}
\hline
Source Name & R.A. & Dec. & Distance To & \textit{WISE} Name & 3.4\,$\mu$m  & 4.6\,$\mu$m & 12\,$\mu$m & 22\,$\mu$m & 850\,$\mu$m & SNR & FIRST  \\
 & (J2000) & (J2000) & \textit{WISE} Target &  (mag) & (mag) & (mag) & (mag) & (mJy) & & &  /NVSS \\
 & & & (arcsec)  & & & & & & & &  Detection \\
& & & & & & & & & & & Limit \\
& & & & & & & & & & &  (mJy/beam) \\
\hline
W0831$+$0140-1 & 08:31:49.565 & 01:41:10.80 & 78$\pm$8 & \textit{WISE} undetected & N/A  & N/A  & N/A  & N/A  & 6.4 $\pm$ 2.1 & 3.0 & 1.00  \\
W0831$+$0140-2 & 08:31:48.498 & 01:40:30.80 & 73$\pm$7 & \textit{WISE} undetected &N/A  & N/A  & N/A  & N/A  &  7.7 $\pm$ 2.1 & 3.7 & 0.99  \\
W0831$+$0140-3 & 08:31:50.899 & 01:39:58.80 & 32$\pm$3 & \textit{WISE} undetected &N/A  & N/A  & N/A  & N/A  &  6.9 $\pm$ 2.1 & 3.3 & 0.99  \\
W1136$+$4236-1 &  11:36:32.868 & 42:35:14.42 & 48$\pm$5 & \textit{WISE} undetected &N/A  & N/A  & N/A  & N/A  & 5.4 $\pm$ 1.7 & 3.2 & 1.06  \\
W1603$+$2747-1$^a$ & 16:03:59.222 & 27:47:05.48 & 70$\pm$7 & \textit{WISE} undetected & N/A  & N/A  & N/A  & N/A  &  6.8 $\pm$ 1.8 & 3.8 & 0.93 \\
W1835$+$4355-1$^a$ & 18:35:28.518 & 43:54:52.36 & 77$\pm$8 & \textit{WISE} undetected &N/A  & N/A  & N/A  & N/A  & 4.6 $\pm$ 1.5 & 3.1 & Undetected    \\
W2216$+$0723-1$^a$  & 22:16:21.520 & 07:24:06.50 & 30$\pm$3 & \textit{WISE} undetected &N/A  & N/A  & N/A  & N/A  &  4.9 $\pm$ 1.6 & 3.1 & 0.80  \\
W2216$+$0723-2$^a$  & 22:16:16.142 & 07:24:34.83 & 33$\pm$3 & \textit{WISE} undetected &N/A  & N/A  & N/A  & N/A  &  5.3 $\pm$ 1.6 & 3.3 & 0.79  \\
W2246$-$0526-1 & 22:46:02.433 & $-$05:26:35.43 & 74$\pm$7 & \textit{WISE} undetected & N/A  & N/A  & N/A  & N/A  &  6.9 $\pm$ 2.1 & 3.3 & 0.97  \\
\hline 
W1814$+$3412-1$^b$ & 18:14:23.130 & 34:12:01.79 & 67$\pm$7 & J181423.13$+$341205.2 &16.848 $\pm$ 0.080 & 16.748  $\pm$0.243 & 12.153  $\pm$ 0.283 & $<$ 8.791 & 5.4 $\pm$ 1.8 & 3.0 & Undetected \\
W2026$+$0716-1 & 20:26:16.913 & 07:15:55.7 & 37$\pm$4 & J202616.55$+$071600.6 &15.336 $\pm$ 0.038 & 15.259 $\pm$ 0.085 & $<$ 12.781 & $<$  9.034 & 7.3 $\pm$ 1.7 & 4.3& Undetected \\
W2026$+$0716-2  & 20:26:20.139 & 07:17:03.57 & 72$\pm$7 & \textit{WISE} undetected&N/A  & N/A  & N/A  & N/A  &  6.1 $\pm$ 1.7 & 3.6 & Undetected   \\
W2054$+$0207-1 & 20:54:22.159 & 02:06:48.38 & 18$\pm$2 & \textit{WISE} undetected &N/A  & N/A  & N/A  & N/A  & 6.9 $\pm$ 1.8 & 3.8& 0.96 \\
W2054$+$0207-2 & 20:54:24.376 & 02:06:46.54 & 16$\pm$2 & \textit{WISE} undetected &N/A  & N/A  & N/A  & N/A  & 5.6 $\pm$ 1.8 & 3.1 & 0.96 \\
W2054$+$0207-3 & 20:54:29.713 & 02:06:19.15 & 68$\pm$7 &  \textit{WISE} undetected &N/A  & N/A  & N/A  & N/A  & 6.8 $\pm$ 1.8 & 3.8 & 0.98\\
W2357$+$0328-1 & 23:57:08.930 & 03:27:11.40 & 48$\pm$5 & J235708.59$+$032712.1 &17.574 $\pm$ 0.212 & 16.679 $\pm$ 0.375 & $<$ 12.031 & $<$ 8.677 & 5.8 $\pm$ 1.9 & 3.1 & 0.86  \\
W2357$+$0328-2 & 23:57:05.457 & 03:27:11.40 & 84$\pm$8 & J235705.39$+$032710.3 &16.536 $\pm$ 0.084 & 16.177 $\pm$ 0.240 & $<$ 11.869 & $<$  8.856 & 6.7 $\pm$ 1.9 & 3.5 & 0.86 \\
\hline
\end{tabular}
\hspace*{-4cm}
\label{ss12fluxes}
\end{table*}
\end{landscape}

\begin{landscape}
\begin{table*}
\caption[Coordinates and photometry of the 81 companion SMG sources found around 30 \textit{WISE}/radio AGNs.]{Coordinates and photometry of the 81 companion SMG sources found around 30 \textit{WISE}/radio AGNs, with 3.4\,$\mu$m, 4.6\,$\mu$m, 12\,$\mu$m and 22\,$\mu$m magnitudes from the All\textit{WISE} Source Catalog and 850\,$\mu$m flux densities from SCUBA-2. The targets with \textit{WISE} upper limits have SNR $<$ 2 and therefore in the All\textit{WISE} Source Catalog the magnitudes quoted are 2\,$\sigma$ upper limits. No \textit{Herschel}, NVSS, FIRST data and no objects found in SIMBAD for all the 81 companion SMG sources. FIRST detection limit for each source position is given in mJy/beam, "undetected" represents that there is no FIRST coverage or NVSS detection. A search radius of 8\,arcsec was used, which is a typical search radius of SMG counterparts \citep{hainline09}.}
\hspace*{-4cm}\begin{tabular}{@{}cccccccccccc@{}}
\hline
Source Name & R.A. & Dec. & Distance To & \textit{WISE} Name  & 3.4\,$\mu$m  & 4.6\,$\mu$m & 12\,$\mu$m & 22\,$\mu$m & 850\,$\mu$m & SNR & FIRST  \\
 & (J2000) & (J2000) & \textit{WISE} Target &  & (mag) & (mag) & (mag) & (mag) & (mJy) & & /NVSS \\
 & & &  (arcsec) & & & & & && & Detection\\
 & & & & & & && & & & Limit \\
& & & & & & & & & & & (mJy/beam) \\
\hline
W0010$+$1643-1  & 00:10:41.983 & 16:43:44.70 & 32$\pm$3 & \textit{WISE} undetected  &N/A  & N/A  & N/A  & N/A  &  6.0 $\pm$ 1.9 & 3.2 & Undetected \\
W0010$+$1643-2  & 00:10:42.284 & 16:43:28.70 & 36$\pm$4 & J001042.69$+$164333.2 &16.349 $\pm$ 0.071 & 16.116 $\pm$ 0.183 & $<$ 11.970 & $<$ 9.008 & 5.9 $\pm$ 1.9 & 3.1 & Undetected \\
W0010$+$1643-3 & 00:10:40.614 & 16:42:52.37 & 34$\pm$4 & J001040.14$+$164253.2  &16.876 $\pm$ 0.108 & 16.268 $\pm$ 0.228 & $<$ 11.943 & $<$ 8.720 & 5.7 $\pm$ 1.9  & 3.0 & Undetected \\
W0010$+$1643-4 & 00:10:36.461 & 16:42:16.37 & 23$\pm$3 & J001036.63$+$164217.1 &17.564 $\pm$ 0.187 & 16.550 $\pm$ 0.295 & $<$ 12.102 & $<$ 8.791 & 6.5 $\pm$ 1.9  &3.4 &  Undetected \\
W0244$+$1123-1 & 02:44:23.184 & 11:24:58.40 & 58$\pm$6 & \textit{WISE} undetected & N/A  & N/A  & N/A  & N/A  & 7.9 $\pm$ 2.1  & 3.8 &  Undetected \\
W0244$+$1123-2 & 02:44:20.191 & 11:24:58.40 & 76$\pm$7 & \textit{WISE} undetected  & N/A  & N/A  & N/A  & N/A  & 6.6 $\pm$ 2.1  & 3.1 & Undetected \\
W0244$+$1123-3 & 02:44:19.421 & 11:24:10.40 & 68$\pm$7 & J024419.14$+$112416.0 9 &14.860 $\pm$ 0.032 & 14.546 $\pm$ 0.051 & $<$ 12.121 & $<$ 9.261 & 6.4 $\pm$ 2.1 & 3.0 & Undetected \\\
W0244$+$1123-4 & 02:44:19.942 & 11:23:10.06 & 74$\pm$7 & \textit{WISE} undetected  &N/A  & N/A  & N/A  & N/A  & 6.9 $\pm$ 2.1  & 3.3 & Undetected \\
W0332$+$3205-1 & 03:32:28.489 & 32:05:28.67 & 21$\pm$3 & J033228.76$+$320525.3  &16.032 $\pm$ 0.061 & 16.060 $\pm$ 0.190 & $<$ 12.561 & $<$ 8.833 & 10.4 $\pm$ 2.0  & 5.2 & Undetected \\ 
W0332$+$3205-2 & 03:32:34.155 & 32:05:48.99 & 69$\pm$7 & \textit{WISE} undetected  &N/A  & N/A  & N/A  & N/A  & 6.5 $\pm$ 2.0  &3.3 &  Undetected \\
W0342$+$3753-1 & 03:42:23.576 & 37:54:37.60 & 60$\pm$6 & J034223.52$+$375442.8 &14.520 $\pm$ 0.030 & 14.523 $\pm$ 0.055 & $<$ 11.797 & $<$ 8.743 & 8.8 $\pm$ 2.1  & 4.2 & Undetected\\
W0342$+$3753-2 & 03:42:22.562 & 37:54:22.93 & 32$\pm$4 & \textit{WISE} undetected  &N/A  & N/A  & N/A  & N/A  & 6.6 $\pm$ 2.1  & 3.1 & Undetected \\
W0342$+$3753-3 & 03:42:20.872 & 37:54:26.27 & 52$\pm$5 & \textit{WISE} undetected  &N/A  & N/A  & N/A  & N/A  & 7.0 $\pm$ 2.1  & 3.3 & Undetected \\
W0342$+$3753-4 & 03:42:16.170 & 37:53:45.92 & 78$\pm$8 & J034216.58$+$375348.5  &17.429 $\pm$ 0.167 & $<$ 16.701 & $<$ 12.640 & $<$ 8.999 & 6.9 $\pm$ 2.1  & 3.3 & Undetected \\
W0342$+$3753-5 & 03:42:28.278 & 37:52:58.26 & 58$\pm$6 & \textit{WISE} undetected  &N/A  & N/A  & N/A  & N/A  & 8.6 $\pm$ 2.1  & 4.1 & Undetected \\
W0342$+$3753-6 & 03:42:28.644 & 37:53:26.26 & 57$\pm$6 & \textit{WISE} undetected  &N/A  & N/A  & N/A  & N/A  & 7.6 $\pm$ 2.1  & 3.6 & Undetected \\
W0342$+$3753-7 & 03:42:20.872 & 37:53:58.27 & 34$\pm$4 & J034221.22$+$375401.9  &15.911 $\pm$ 0.056 & 16.036 $\pm$ 0.167 & $<$ 11.844 & $<$ 8.938  & 7.4 $\pm$ 2.1 & 3.5 & Undetected \\
W0352$+$1947-1 & 03:52:09.368 & 19:47:08.86 & 54$\pm$5 & \textit{WISE} undetected  &N/A  & N/A  & N/A  & N/A  &  6.0 $\pm$ 1.9 & 3.2 & Undetected \\
W0404$+$0712-1 & 04:04:41.751 & 07:12:59.20 & 40$\pm$4 & J040441.89$+$071258.0 &17.185 $\pm$ 0.145 & $<$ 17.301 & $<$ 12.608 & $<$ 9.134 & 6.2 $\pm$ 1.9  & 3.3 & Undetected \\
W0404$+$0712-2 & 04:04:43.566 & 07:12:30.20 & 35$\pm$4 & \textit{WISE} undetected  &N/A  & N/A  & N/A  & N/A  & 6.1 $\pm$ 1.9  &3.2 &  Undetected \\
W0443$+$0643-1 & 04:43:32.581 & 06:43:50.10 & 32$\pm$4 & \textit{WISE} undetected  &N/A  & N/A  & N/A  & N/A  & 13.5 $\pm$ 3.7  & 3.6 & Undetected \\
W0443$+$0643-2 & 04:43:34.230 & 06:42:09.81 & 68$\pm$7 & J044334.58$+$064212.9 &15.757 $\pm$ 0.049 & 15.578 $\pm$ 0.140 & $<$ 12.240 & $<$ 8.843 & 12.0 $\pm$ 3.7  & 3.2 & Undetected \\
W0443$+$0643-3 & 04:43:35.534 & 06:42:18.67 & 70$\pm$7 & \textit{WISE} undetected  &N/A  & N/A  & N/A  & N/A  &  6.0 $\pm$ 1.9 & 3.2 & Undetected \\
W0849$+$3033-1 & 08:49:02.961 & 30:32:33.00 & 62$\pm$6 & \textit{WISE} undetected  &N/A  & N/A  & N/A  & N/A  & 7.7 $\pm$ 2.4 & 3.2 & 0.93 \\
\hline
\end{tabular}
\hspace*{-4cm}
\label{ss13fluxes}
\end{table*}
\end{landscape}

\begin{landscape}
\begin{table*}
\caption[Continue table of coordinates and photometry of the 81 companion SMG sources found around 30 \textit{WISE}/radio AGNs.]{Continue table of coordinates and photometry of the 81 companion SMG sources found around 30 \textit{WISE}/radio AGNs, with 3.4\,$\mu$m, 4.6\,$\mu$m, 12\,$\mu$m and 22\,$\mu$m magnitudes from the All\textit{WISE} Source Catalog and 850\,$\mu$m flux densities from SCUBA-2. The targets with \textit{WISE} upper limits have SNR $<$ 2 and therefore in the All\textit{WISE} Source Catalog the magnitudes quoted are 2\,$\sigma$ upper limits. No \textit{Herschel}, NVSS, FIRST data and no objects found in SIMBAD for all the 81 companion SMG sources. FIRST detection limit for each source position is given in mJy/beam, "undetected" represents that there is no FIRST coverage or NVSS detection. A search radius of 8\,arcsec was used, which is a typical search radius of SMG counterparts \citep{hainline09}.}
\hspace*{-4cm}\begin{tabular}{@{}cccccccccccc@{}}
\hline
Source Name & R.A. & Dec. & Distance To & \textit{WISE} Name  & 3.4\,$\mu$m  & 4.6\,$\mu$m & 12\,$\mu$m & 22\,$\mu$m & 850\,$\mu$m & SNR & FIRST  \\
 & (J2000) & (J2000) & \textit{WISE} Target &  & (mag) & (mag) & (mag) & (mag) & (mJy) & & /NVSS \\
 & & & (arcsec) & & & & & & && Detection\\
 & & & & & & && & & & Limit \\
& & & & & & &  & & & &(mJy/beam) \\
\hline
W0849$+$3033-2 & 08:49:05.438 & 30:32:12.33 & 76$\pm$8 & \textit{WISE} undetected  & N/A  & N/A  & N/A  & N/A  & 7.5 $\pm$ 2.4 & 3.1& 0.93 \\
W0849$+$3033-3 & 08:49:07.270 & 30:32:12.33 & 83$\pm$8 & \textit{WISE} undetected  & N/A  & N/A  & N/A  & N/A  & 7.9 $\pm$ 2.4 &3.3 & 0.92 \\
W1025$+$6128-1 & 10:25:02.081 & 61:28:49.02 & 54$\pm$5 & \textit{WISE} undetected  & N/A  & N/A  & N/A  & N/A  & 6.3 $\pm$ 2.0 & 3.2 &0.98 \\
W1025$+$6128-2 & 10:25:12.691 & 61:27:40.36 & 58$\pm$6& \textit{WISE} undetected  & N/A  & N/A  & N/A  & N/A  & 6.0 $\pm$ 2.0 & 3.0 & 0.98 \\
W1025$+$6128-3 & 10:25:14.969 & 61:27:21.36 & 78$\pm$8 & \textit{WISE} undetected  & N/A  & N/A  & N/A  & N/A  & 6.1 $\pm$ 2.0 & 3.1 & 0.98 \\
W1046$-$0250-1 &10:46:31.465 & $-$02:50:06.70 & 30$\pm$4 & J104631.08$-$025001.8 &17.645 $\pm$ 0.216 & 16.490 $\pm$  0.283 & $<$ 12.532 & $<$ 8.979 & 6.7 $\pm$ 2.1 & 3.2 & 0.90 \\
W1046$-$0250-2 &10:46:30.931 & $-$02:49:11.03 & 75$\pm$8 & \textit{WISE} undetected &N/A  & N/A  & N/A  & N/A  & 7.2 $\pm$ 2.1 & 3.4 & 0.90 \\
W1046$-$0250-3 &10:46:37.272 & $-$02:50:32.70 & 59$\pm$6 & \textit{WISE} undetected &N/A  & N/A  & N/A  & N/A  & 6.2 $\pm$ 2.1 & 3.0 & 0.90 \\
W1107$+$3421-1 & 11:07:32.012 & 34:20:55.03 & 34$\pm$4 & \textit{WISE} undetected  & N/A  & N/A  & N/A  & N/A  & 7.0 $\pm$ 2.0 & 3.0 & 0.89 \\
W1107$+$3421-2  &11:07:30.425 & 34:19:51.36 & 88$\pm$8 & \textit{WISE} undetected    &N/A  & N/A  & N/A  & N/A  & 6.4 $\pm$ 2.0 & 3.2 & 0.90 \\
W1210$+$4750-1 & 12:10:30.748 & 47:50:51.20 & 49$\pm$5 & J121031.00$+$475052.0& 17.411 $\pm$ 0.141 & 16.638 $\pm$ 0.254 & $<$ 12.681 & $<$ 8.527 & 8.2 $\pm$ 2.3 & 3.6 & 0.96 \\
W1210$+$4750-2 & 12:10:37.038 & 47:50:30.51 & 88$\pm$8 & \textit{WISE} undetected  & N/A  & N/A  & N/A  & N/A  & 7.0 $\pm$ 2.3 & 3.0 & 0.96 \\
W1210$+$4750-3 & 12:10:30.515 & 47:49:50.86 & 28$\pm$3 & J121030.53$+$474956.7 & 16.952 $\pm$ 0.095 & 16.452 $\pm$ 0.203 & $<$ 12.641 & $<$ 9.071 & 7.6 $\pm$ 2.3 &3.3 &  0.97 \\
W1210$+$4750-4 & 12:10:26.676 & 47:48:38.87 & 85$\pm$8 & J121027.36$+$474837.9 & 17.364 $\pm$ 0.129 & 17.296 $\pm$ 0.468 & $<$ 12.331 & $<$ 8.509 & 10.7 $\pm$ 2.3 & 4.7 & 0.95 \\
W1210$+$4750-5 & 12:10:21.181 & 47:49:27.85 & 75$\pm$8 & J121021.04$+$474927.4 & 17.210 $\pm$ 0.124 & 16.402 $\pm$ 0.207 & $<$ 12.546 & $<$ 8.992 & 7.7 $\pm$ 2.3 & 3.3 & 0.99\\
W1212$+$4659-1 & 12:12:08.419 & 47:00:07.23 & 36$\pm$4 & J121208.25$+$470004.4& 16.950 $\pm$ 0.103 & 16.375 $\pm$ 0.211 & $<$ 12.261 & $<$ 8.896 & 7.9 $\pm$ 2.5 & 3.2 & 0.99 \\
W1212$+$4659-2 & 12:12:10.015 & 46:59:34.56 & 48$\pm$5 & \textit{WISE} undetected  & N/A  & N/A  & N/A  & N/A  &  6.0 $\pm$ 1.9 & 3.2 & 0.98 \\ 
W1212$+$4659-3 & 12:12:11.970 & 46:59:42.22 & 63$\pm$6 & \textit{WISE} undetected  & N/A  & N/A  & N/A  & N/A  & 7.8 $\pm$ 2.5 & 3.1 & 1.58 \\
W1409$+$1732-1 & 14:09:22.400 & 17:32:20.03 & 82 & \textit{WISE} undetected  & N/A  & N/A  & N/A  & N/A  & 6.4 $\pm$ 2.0 & 3.2 & 1.58 \\
W1409$+$1732-2 & 14:09:26.832 & 17:31:44.35 & 30$\pm$4 & J140927.37$+$173142.7 & 17.603 $\pm$ 0.174 & $<$ 16.764 & 12.961 $\pm$ 0.526 & $<$ 8.759 & 6.4 $\pm$ 2.0 & 3.2& 1.58 \\ 
W1409$+$1732-3 & 14:09:27.950 & 17:31:07.73 & 59$\pm$6 & J140928.23$+$173108.2 & 17.956 $\pm$ 0.211 & 16.908 $\pm$ 0.311 & $<$ 12.657 & $<$ 9.339 & 7.9 $\pm$ 2.0 & 4.0 & 0.93 \\ 
W1428$+$1113-1 & 14:29:00.256 & 17:12:11.25 & 68$\pm$7 & \textit{WISE} undetected   & N/A  & N/A  & N/A  & N/A  &  6.0 $\pm$ 1.9 & 3.7 & 1.0 \\ 
W1501$+$1324-1 & 15:01:41.713 & 13:24:57.90 & 44$\pm$4 & J150141.93$+$132450.8 & 18.011 $\pm$ 0.214 & $<$ 17.207 & $<$ 12.892 & $<$ 9.321 & 7.1 $\pm$ 2.2 & 3.2 & 0.95 \\
W1501$+$1324-2 & 15:01:42.192 & 13:23:57.23 & 63$\pm$6 & J150142.45$+$132358.6 & 17.196 $\pm$ 0.106 & 16.988 $\pm$ 0.330 & $<$ 12.984 & $<$ 9.239 & 7.2 $\pm$ 2.2 & 3.3 & 0.94 \\
\hline
\end{tabular}
\hspace*{-4cm}
\label{ss13fluxes2}
\end{table*}
\end{landscape}

\begin{landscape}
\begin{table*}
\caption[Continue table of coordinates and photometry of the 81 companion SMG sources found around 30 \textit{WISE}/radio AGNs.]{Continue table of coordinates and photometry of the 81 companion SMG sources found around 30 \textit{WISE}/radio AGNs, with 3.4\,$\mu$m, 4.6\,$\mu$m, 12\,$\mu$m and 22\,$\mu$m magnitudes from the All\textit{WISE} Source Catalog and 850\,$\mu$m flux densities from SCUBA-2. The targets with \textit{WISE} upper limits have SNR $<$ 2 and therefore in the All\textit{WISE} Source Catalog the magnitudes quoted are 2\,$\sigma$ upper limits. No \textit{Herschel}, NVSS, FIRST data and no objects found in SIMBAD for all the 81 companion SMG sources. FIRST detection limit for each source position is given in mJy/beam, "undetected" represents that there is no FIRST coverage or NVSS detection. A search radius of 8\,arcsec was used, which is a typical search radius of SMG counterparts \citep{hainline09}.}
\hspace*{-4cm}\begin{tabular}{@{}cccccccccccc@{}}
\hline
Source Name & R.A. & Dec. & Distance To & \textit{WISE} Name  & 3.4\,$\mu$m  & 4.6\,$\mu$m & 12\,$\mu$m & 22\,$\mu$m & 850\,$\mu$m & SNR & FIRST  \\
 & (J2000) & (J2000) & \textit{WISE} Target &  & (mag) & (mag) & (mag) & (mag) & (mJy) & & /NVSS \\
 & & & (arcsec) & & & && & & & Detection\\
 & & & & & & && & & & Limit \\
& & & & & & && & & &  (mJy/beam) \\
\hline
W1501$+$1324-3 & 15:01:40.022 & 13:23:33.90 & 76$\pm$7 & \textit{WISE} undetected &N/A  & N/A  & N/A  & N/A  & 7.3 $\pm$ 2.2 & 3.3 & 0.94 \\
W1501$+$1324-4 & 15:01:38.423 & 13:24:33.57 & 20$\pm$3 & \textit{WISE} undetected &N/A  & N/A  & N/A  & N/A  & 6.8 $\pm$ 2.2  & 3.1 & 0.94 \\
W1517$+$3523-1 & 15:17:59.282 & 35:24:29.97 & 32$\pm$3 & \textit{WISE} undetected &N/A  & N/A  & N/A  & N/A  & 5.8 $\pm$ 1.9  & 3.1 & 0.94 \\
W1517$+$3523-2 & 15:17:56.992 & 35:23:49.97 & 24$\pm$3 & \textit{WISE} undetected &N/A  & N/A  & N/A  & N/A  & 5.7 $\pm$ 1.9  & 3.0 & 0.99 \\
W1517$+$3523-3 & 15:17:51.703 & 35:23:33.95 & 88$\pm$8 & J151751.39$+$352327.9 &17.874 $\pm$ 0.158 & $<$ 17.702 & $<$ 12.762 & $<$ 9.464 & 5.7 $\pm$ 1.9  & 3.0 & 0.99 \\
W1517$+$3523-4 & 15:17:54.348 & 35:23:02.30 & 72$\pm$7 & J151754.19$+$352302.8 &16.418 $\pm$ 0.052 & 15.611 $\pm$ 0.084 & 12.607 $\pm$ 0.319 & $<$ 9.495 & 5.7 $\pm$ 1.9  & 3.0 & 0.99 \\
W1517$+$3523-5 & 15:17:56.065 & 35:22:35.30 & 74$\pm$7 & \textit{WISE} undetected &N/A  & N/A  & N/A  & N/A  & 5.8 $\pm$ 1.9  &3.1 & 0.99 \\
W1630$+$5126-1 & 16:30:41.014 & 51:26:52.03 & 56$\pm$6 & \textit{WISE} undetected &N/A  & N/A  & N/A  & N/A  & 5.8 $\pm$ 1.9  &3.1 & 0.96 \\
W1630$+$5126-2 & 16:30:30.174 & 51:27:04.36 & 73$\pm$7 & \textit{WISE} undetected &N/A  & N/A  & N/A  & N/A  & 5.8 $\pm$ 1.9  &3.1 & 0.97 \\
W1703$+$2615-1 & 17:03:35.673 & 26:16:19.28 & 70$\pm$7 & \textit{WISE} undetected &N/A  & N/A  & N/A  & N/A  & 6.8 $\pm$ 2.0  & 3.4 & 0.98 \\
W1717$+$5313-1 & 17:17:00.517 & 53:13:51.29 & 49$\pm$5 & \textit{WISE} undetected &N/A  & N/A  & N/A  & N/A  & 7.5 $\pm$ 2.5  & 3.0 & 0.95 \\
W1717$+$5313-2 & 17:17:00.592 & 53:13:15.63 & 53$\pm$5 & J171700.48$+$531319.7&15.618 $\pm$ 0.341 & 15.988 $\pm$ 0.195 & $<$ 10.482 & $<$ 8.604 & 11.7 $\pm$ 2.5  & 4.7 & 0.95 \\
W1717$+$5313-3 & 17:17:08.461 & 53:13:15.96 & 40$\pm$4 & \textit{WISE} undetected &N/A  & N/A  & N/A  & N/A  & 7.5 $\pm$ 2.5  & 3.0 & 0.96 \\
W2126$-$0103-1  &21:26:17.587 & $-$01:04:18.48 & 50$\pm$5 & \textit{WISE} undetected &N/A  & N/A  & N/A  & N/A  & 7.8 $\pm$ 2.0  & 3.9 & 1.0  \\
W2133$-$1419-1  &21:33:53.760 & $-$14:19:46.42 & 58$\pm$6 & J213353.65$-$141946.5 &15.002 $\pm$ 0.037 & 14.929 $\pm$ 0.080 & $<$ 12.506 & $<$ 8.595 & 5.5 $\pm$ 1.8 & 3.1 & Undetected \\
W2212$+$3326-1 & 22:12:54.092 & 33:26:36.35 & 63$\pm$6 & \textit{WISE} undetected &N/A  & N/A  & N/A  & N/A  & 8.9 $\pm$ 2.0 & 4.5 & Undetected\\
W2212$+$3326-2 & 22:13:01.443 & 33:27:24.72 & 86$\pm$8 & \textit{WISE} undetected &N/A  & N/A  & N/A  & N/A  & 6.7 $\pm$ 2.0 & 3.4 & Undetected\\
W2212$-$1253-1  &22:12:00.655 & $-$12:53:38.03 & 59$\pm$6 & \textit{WISE} undetected &N/A  & N/A  & N/A  & N/A  & 8.0 $\pm$ 2.1 & 3.8 & Undetected \\
W2212$-$1253-2  &22:12:09.226 & $-$12:54:07.70 & 59$\pm$6 & \textit{WISE} undetected &N/A  & N/A  & N/A  & N/A  & 7.5 $\pm$ 2.1 & 3.6 & Undetected \\
W2222$+$0951-1 & 22:22:44.211 & 09:51:18.11 & 58$\pm$6 & \textit{WISE} undetected &N/A  & N/A  & N/A  & N/A  & 7.3 $\pm$ 2.0  &3.7 &  0.86\\
W2222$+$0951-2 & 22:22:46.971 & 09:50:09.38 & 78$\pm$7 & J222247.42$+$095008.1 &17.476 $\pm$ 0.180 & $<$ 17.290 & $<$ 12.598 & $<$ 8.629 & 9.3 $\pm$ 2.0  &4.7 &  0.85\\
W2222$+$0951-3 & 22:22:50.953 & 09:50:29.02 & 62$\pm$6 & \textit{WISE} undetected &N/A  & N/A  & N/A  & N/A  & 8.6 $\pm$ 2.0  & 4.3 & 0.86\\
W2226$+$0025-1 & 22:26:20.255 & 00:24:35.57 & 69$\pm$7 & \textit{WISE} undetected &N/A  & N/A  & N/A  & N/A  & 8.4 $\pm$ 1.9  & 4.4 & 0.77\\
W2226$+$0025-2 & 22:26:21.855 & 00:24:04.23 & 74$\pm$7 & \textit{WISE} undetected &N/A  & N/A  & N/A  & N/A  & 6.9 $\pm$ 1.9  & 3.7 & 0.77\\ 
\hline
\end{tabular}
\hspace*{-4cm}
\label{ss13fluxes4}
\end{table*}
\end{landscape}

\begin{landscape}
\begin{table*}
\caption[Continue table of coordinates and photometry of the 81 companion SMG sources found around 30 \textit{WISE}/radio AGNs.]{Continue table of coordinates and photometry of the 81 companion SMG sources found around 30 \textit{WISE}/radio AGNs, with 3.4\,$\mu$m, 4.6\,$\mu$m, 12\,$\mu$m and 22\,$\mu$m magnitudes from the All\textit{WISE} Source Catalog and 850\,$\mu$m flux densities from SCUBA-2. The targets with \textit{WISE} upper limits have SNR $<$ 2 and therefore in the All\textit{WISE} Source Catalog the magnitudes quoted are 2\,$\sigma$ upper limits. No \textit{Herschel}, NVSS, FIRST data and no objects found in SIMBAD for all the 81 companion SMG sources. FIRST detection limit for each source position is given in mJy/beam, "undetected" represents that there is no FIRST coverage or NVSS detection. A search radius of 8\,arcsec was used, which is a typical search radius of SMG counterparts \citep{hainline09}.}
\hspace*{-4cm}\begin{tabular}{@{}cccccccccccc@{}}
\hline
Source Name & R.A. & Dec. & Distance To & \textit{WISE} Name  & 3.4\,$\mu$m  & 4.6\,$\mu$m & 12\,$\mu$m & 22\,$\mu$m & 850\,$\mu$m & SNR & FIRST  \\
 & (J2000) & (J2000) & \textit{WISE} Target &  & (mag) & (mag) & (mag) & (mag) & (mJy) & & /NVSS \\
 & & & (arcsec) & & & & && & & Detection\\
 & & & & & & & & & & &Limit \\
& & & & & & & & & & & (mJy/beam) \\
\hline
W2230$-$0720-1 & 22:30:06.993 & $-$07:19:51.90 & 70$\pm$7 & \textit{WISE} undetected  &N/A  & N/A  & N/A  & N/A  &  5.8 $\pm$ 1.8 & 3.2 & 0.93\\
W2230$-$0720-2 & 22:30:09.189& $-$07:19:59.90 & 57$\pm$6 & \textit{WISE} undetected &N/A  & N/A  & N/A  & N/A  & 6.0 $\pm$ 1.8 & 3.3 & 0.93\\
W2230$-$0720-3 & 22:30:12.080& $-$07:20:47.90 & 59$\pm$6 & \textit{WISE} undetected  &N/A  & N/A  & N/A  & N/A  &  5.5 $\pm$ 1.8 & 3.1 & 0.93\\
W2230$-$0720-4 & 22:30:08.607& $-$07:22:24.23 & 80$\pm$8 & J223008.35$-$072220.9 &17.423 $\pm$ 0.174 & 16.992 $\pm$ 0.438 & $<$ 12.065 & $<$ 8.834 & 5.7 $\pm$ 1.8 & 3.2 & 0.92 \\
W2325$-$0429-1 & 23:25:06.972& $-$04:28:44.43 & 66$\pm$7 & \textit{WISE} undetected &N/A  & N/A  & N/A  & N/A  & 7.1 $\pm$ 2.1 & 3.4 & 2.27\\
W2325$-$0429-2 & 23:25:01.088& $-$04:30:43.77 & 76$\pm$7 & \textit{WISE} undetected  &N/A  & N/A  & N/A  & N/A  &  9.8 $\pm$ 2.1 & 4.7 & 2.20\\
W2331$-$1411-1 & 23:31:05.148& $-$14:11:08.87 & 48$\pm$5 & J233104.89$-$141108.4 & 16.393 $\pm$ 0.074 & 16.900 $\pm$ 0.423 & $<$ 12.543 & $<$ 8.351 & 6.9 $\pm$ 2.2 & 3.1 & Undetected\\
W2345$+$3120-1 & 23:45:44.634& 31:19:41.80 & 56$\pm$6 & J234544.45$+$311941.9   &17.217 $\pm$ 0.124 & 15.988 $\pm$ 0.157 & $<$ 12.585 & $<$ 8.841 & 11.3 $\pm$ 2.0 & 5.7 & Undetected\\
W2345$+$3120-2 & 23:45:34.358& 31:20:02.12 & 84$\pm$8 & \textit{WISE} undetected &N/A  & N/A  & N/A  & N/A  & 6.3 $\pm$ 2.0 & 3.2 & Undetected\\
\hline
\end{tabular}
\hspace*{-4cm}
\label{ss13fluxes6}
\end{table*}
\end{landscape}

\begin{table*}
\caption[SFRDs of Hot DOGs and \textit{WISE}/radio AGNs.]{SFRDs of Hot DOGs (top 10) and \textit{WISE}/radio AGNs (bottom 30) and their surrounding SMGs, assuming they are at the same redshift. The angular radius is estimated to be the size of the SCUBA-2 map, 1.5\,arcmin. The spectroscopically known redshifts of the \textit{WISE}-selected targets are shown. The unknown redshifts of the \textit{WISE}/radio AGNs are assumed to the average of \textit{WISE}/radio AGNs, $ z = 1.7 $. }
\begin{tabular}{@{}ccc@{}}
\hline
Source & SFRD &Redshift \\
 & M$_{\odot}$ yr$^{-1}$ Mpc$^{-3}$ &  \\
\hline
W0831$+$0140 & 7949$\pm$159   & 3.91 \\
W1136$+$4236 & 2064$\pm$41  & 2.39\\
W1603$+$2747 & 2989$\pm$60  & 2.63 \\
W1814$+$3412 & 1523$\pm$30  & 2.45 \\
W1835$+$4355 & 2406$\pm$48  & 2.3 \\
W2026$+$0716 & 2835$\pm$57  & 2.54 \\
W2054$+$0207 & 4304$\pm$86  & 2.52  \\
W2216$+$0723 & 2917$\pm$58  & 1.68 \\
W2246$-$0526 & 5808$\pm$116  & 4.59 \\
W2357$+$0328 & 2538$\pm$51  & 2.12 \\
\hline
W0010$+$1643 & 5311$\pm$106  & 2.855 \\
W0244$+$1123 & 4035$\pm$162  & Unknown \\
W0332$+$3205 & 1799$\pm$72  & Unknown\\
W0342$+$3753 & 18715$\pm$374  & 0.47 \\
W0352$+$1947 & 1045$\pm$42  & Unknown\\
W0404$+$0712 & 2133$\pm$85  & Unknown\\
W0443$+$0643 & 3686$\pm$147  & Unknown\\
W0849$+$3033 & 4005$\pm$160  & Unknown\\
W1025$+$6128 & 3091$\pm$124  & Unknown\\
W1046$-$0250 & 2699$\pm$108  & Unknown\\
W1107$+$3421 & 2119$\pm$42  & Unknown\\
W1210$+$4750 & 4992$\pm$100 & Unknown \\
W1212$+$4659 & 3178$\pm$64 & Unknown \\
W1409$+$1732 & 3454$\pm$138  & Unknown\\
W1428$+$1113 & 1219$\pm$49  & 1.6\\
W1501$+$1324 & 9501$\pm$190  & 0.505 \\
W1517$+$3523 & 4499$\pm$90  & 1.515 \\
W1630$+$5126 & 2046$\pm$82  & Unknown\\
W1703$+$2615 & 1364$\pm$55  & Unknown\\
W1717$+$5313 & 3935$\pm$157  & 2.717 \\
W2126$-$0103 & 2247$\pm$89  & 0.607 \\
W2133$-$1419 & 1132$\pm$45  & Unknown\\
W2212$-$1253 & 2931$\pm$117  & Unknown\\
W2212$+$3326 & 1799$\pm$72  & Unknown\\
W2222$+$0951 & 3149$\pm$63  & Unknown\\
W2226$+$0025 & 4025$\pm$80  & 0.607 \\
W2230$-$0720 & 12596$\pm$252  & 0.444 \\
W2325$-$0429 & 2220$\pm$88  & 1.737 \\
W2331$-$1411 & 2177$\pm$87 & Unknown \\
W2345$+$3120 & 2772$\pm$111  & Unknown\\
\hline
\end{tabular}
\label{sfrd}
\end{table*}

\label{lastpage}
\end{document}